\pdfoutput=1
\documentclass[a4paper, 11pt]{article}
\usepackage{geometry}
\usepackage[pdftex]{graphicx,color} 
\usepackage{jheppub}
\usepackage{amsmath}
\usepackage{amssymb}
\usepackage{comment}
\usepackage{multirow}
\usepackage{mathtools}
\usepackage[boxsize=1em,centertableaux]{ytableau}
\usepackage{dsdshorthand}
\usepackage{braket}
\usepackage{tocloft}
\usepackage{fnpct}
\usepackage{slashed}
\usepackage{upgreek}
\usepackage{dsfont}

\usepackage{tikz}
\usepackage{subcaption}
\usepackage{float}

\usetikzlibrary{arrows,calc,shapes,decorations.pathmorphing,decorations.markings,positioning}
\tikzset{
>=stealth',
help lines/.style={dashed, thick},
axis/.style={<->},
important line/.style={thick},
connection/.style={thick, dotted},
  cross/.style={
    cross out,
    draw=black, 
    minimum size=7pt, 
    inner sep=0pt,
    outer sep=0pt
  },
  branchcut/.style={
    decoration={
      snake,
      amplitude=1pt,
      segment length=6pt,
    },
    decorate,
    thick
  },
->-/.style={decoration={
  markings,
  mark=at position #1 with {\arrow{>}}},postaction={decorate}},
  twopt/.style={
    circle,
    draw,
    fill=black,
    inner sep=1pt,
    minimum size=1pt
  },
  scalar/.style={
    thick,
    dashed,
    postaction={
      decorate,
      decoration={
        markings,
        mark=at position 0.5 with {\arrow{>}}
      }
    }
  },
  spinning/.style={
    thick,
    postaction={
      decorate,
      decoration={
        markings,
        mark=at position 0.5 with {\arrow{>}}
      }
    }
  },
  scalar no arrow/.style={
    thick,
    dashed,
  },
  spinning no arrow/.style={
    thick,
  },
  finite/.style={
    decoration={
      snake,
      amplitude=1pt,
      segment length=6pt,
    },
    decorate,
    thick
  },
  axis/.style={
    thick,
    postaction={
      decorate,
      decoration={
        markings,
        mark=at position 1 with {\arrow{>}}
      }
    }
  },
}

\tikzset{snake it/.style={decorate, decoration=snake}}

\newcommand{\bea}{\begin{equation}\begin{aligned}}
\newcommand{\eea}[1]{\label{#1}\end{aligned}\end{equation}}
\newcommand{\beq}{\begin{equation}}
\newcommand{\eeq}{\end{equation}}

\def\1{{\rm 1-loop}}

\newcommand{\xdownarrow}[1]{%
  {\left\downarrow\vbox to #1{}\right.\kern-\nulldelimiterspace}
}

\newcommand{\tens}[1]{%
  \mathbin{\mathop{\otimes}\limits_{#1}}%
}

\newcommand{\tensadd}[1]{%
  \mathbin{\mathop{\oplus}\limits_{#1}}%
}

\makeatletter
\def\mathcolor#1#{\@mathcolor{#1}}
\def\@mathcolor#1#2#3{%
  \protect\leavevmode
  \begingroup\color#1{#2}#3\endgroup
}
\makeatother

\title{A constructive solution to the cosmological bootstrap.}
\author{Agnese Bissi$ ^{\dagger \Diamond}$ Sourav Sarkar$ ^{\dagger}$}
\affiliation{$ ^{\dagger}$Department of Physics and Astronomy,\\
Uppsala University, Box 516, SE-75120, Uppsala, Sweden.}
\affiliation{$ ^{\Diamond}$ICTP, International Centre for Theoretical Physics\\
Strada Costiera 11, 34151, Trieste, Italy.
}
\emailAdd{agnese.bissi@physics.uu.se, sourav.sarkar@physics.uu.se}
\keywords{CFT, AdS/CFT, Conformal bootstrap, dS bootstrap}

\abstract{In this paper we revisit a generalised crossing equation that follows from harmonic analysis on the conformal group, and is of particular interest for the cosmological bootstrap programme. We present an exact solution to this equation, for dimensions two or higher, in terms of 6j symbols of the Euclidean conformal group, and discuss its relevance. In the process we provide a detailed derivation of the analogue of the Biedenharn-Elliot identity for said 6j symbols. }

\begin{document}
\maketitle

%!TEX root = ../Central_compile.tex
%%%%%%%%%%%%%%%%%%%%%%%%%%%%%%%%%%%%%%%%%%%%%

\section{Introduction}
\label{sec:intro}
Conformal symmetry strongly constrains the space-time structure of observables in conformal field theories (CFTs). This feature, together with the existence of an Operator Product Expansion (OPE), has made it possible to classify and study CFTs using their symmetries.

%In this section, we shall pose the problem that is the position independent crossing equation, to which we wish to find a solution. On the way we shall first briefly review some (Euclidean) conformal representation theory, and how conformal correlators can be expanded in a complete basis of conformal partial waves (CPW). A (position dependent) crossing equation using CPWs follows directly from this completeness property and the position independent crossing equation can then be obtained using the orthonormality of CPWs \cite{Gadde:2017sjg, Hogervorst:2021uvp}. An additional unitarity constraint that can be imposed on a tentative solution follows from the unitarity of the associated representations. Thereafter we shall briefly discuss the role of this crossing equation in the context of the cosmological bootstrap \cite{Hogervorst:2021uvp}. 

When studying CFTs that are physically relevant, we are typically interested in unitary Lorentzian theories. Their symmetry group is $SO(d,2)$, and the Unitary Irreducible Representations (UIRs) are different from those of $SO(d+1,1)$. By virtue of the Osterwalder-Schraeder reconstruction theorem \cite{cmp/1103858969}, we can generally obtain a unitary Lorentzian theory by analytically continuing a reflection positive Euclidean theory. Reflection positivity imposes the familiar unitarity bounds on CFT data \cite{Mack1977,Jantzen1977,Minwalla:1997ka},
\bea
\Delta \geq \left\{ \begin{array}{ccc} \frac{d-2}{2} & \quad \text{for} & \ell = 0 , \\  \ell +d -2  & \quad \text{for} & \ell > 0 , \end{array} \right.
\eea{unitaritybounds}
in addition to the unit operator for the disconnected piece in the correlators.

The conformal bootstrap programme \cite{FERRARA1973161,Polyakov:1974gs,Rattazzi:2008pe} aims to constrain the set of permissible values of $[\Delta, \ell]$ based on general constraints like \eqref{unitaritybounds}  and the associativity of the OPE, without any reference to a microscopic description of the theory. The most important constraint that is used is the crossing equation, a consequence of the associativity of the OPE in a four-point correlator\footnote{We explain our conventions for the three-point function coefficients and tensor structures in App. \ref{sec:tsconv}},
\beq
\braket{O_1 (x_1) O_2 (x_2) O_3 (x_3) O_4 (x_4)} = \sum_{\Delta_E ,\ell_E } \; \lambda_{12E}\lambda_{34E} \; G^{E}_{1234} (x_i )   = \sum_{\Delta_E' ,\ell_E' } \; \lambda_{14E'}\lambda_{32E'} \; G^{E'}_{3214} (x_i ) .      \label{cbcrossing}
\eeq 
This equation comes from equating the conformal block expansion of the correlator in two different channels\footnote{Note that the channel is indicated by the labels on the three point function coefficients, and the labels in subscript of the block. $G_{ijkl}$ is the conformal block in the channel $(ij)-(kl)$.}. Four-point correlation functions in a unitary Lorentzian CFT can be expanded in conformal blocks with quantum numbers satisfying the bounds in \eqref{unitaritybounds}\footnote{It is customary to use the term ``conformal block" for a conformal invariant function of the cross-ratios that solves the Casimir equation (expressed in cross-ratios) and has a boundary behaviour consistent with the OPE. We however include the conformally covariant factors in our usage of the term.}. This expansion is convergent (in an appropriate domain) owing to the fact that the Operator Product Expansion (OPE) in a CFT is convergent \cite{Ferrara:1971zy,Ferrara:1971vh,Ferrara:1972cq,Mack:1976pa,Pappadopulo:2012jk,Rychkov:2015lca}. $\lambda_{12E}\lambda_{34E}$ are OPE coefficients and they depend on the dynamics of the specific theory. In a unitary CFT, they are real and hence the product is $\geq 0$ for pairwise identical operators. Note that conformal blocks are not single-valued functions of position variables and have no orthogonality properties. The expansion in \eqref{cbcrossing} is not a completeness relation and the quantum numbers $[\Delta_E ,\ell_E]$ appearing in the sum are theory dependent. 

Due to these properties of the conformal blocks, finding solutions to the crossing relations \eqref{cbcrossing} for non trivial CFTs in $d \ge 2 $ is a complicated task.  In recent years, there has been a lot of progress in this direction stemming from the conformal bootstrap programme.  The numerical bootstrap programme \cite{Rattazzi:2008pe,Poland:2018epd,Poland:2022qrs} uses the crossing equation \eqref{cbcrossing} and unitarity together with convex optimization techniques to impose bounds on the conformal dimensions and OPE coefficients for different classes of strongly interacting theories. Intriguingly,  some physically interesting theories are close to saturating these bounds. The prototypical example is the three dimensional Ising model, for which  this has led us to precise values for the critical exponents  \cite{PhysRevD.86.025022, El-Showk:2014dwa, Kos:2014bka, Simmons-Duffin:2015qma, Kos:2016ysd}. On the other hand, analytical bootstrap \cite{Hartman:2022zik} has led us to new structures in generic CFT spectra, for example the existence of a large spin sector \cite{Alday:2007mf,Fitzpatrick:2012yx,Komargodski:2012ek} and multi-twist operators, which in turn follows from the organization of local operators in Regge trajectories \cite{Costa:2012cb} that are analytical in spin \cite{Caron_Huot_2017,Simmons_Duffin_2018}. Conformal bootstrap also gives us tools to constrain quantum gravity in AdS \cite{Caron-Huot:2021enk, Afkhami-Jeddi:2016ntf, Kulaxizi:2017ixa, Costa:2017twz, Meltzer:2017rtf} and in particular, provides access to weakly coupled AdS gravitational theories without having the need to use perturbation theory in AdS \cite{Aharony:2016dwx, Heemskerk:2009pn, Fitzpatrick:2010zm, Meltzer:2020qbr,Meltzer:2019nbs}\footnote{See \cite{Poland:2022qrs,Hartman:2022zik,Bissi:2022mrs,Kruczenski:2022lot,Gopakumar:2022kof} for a detailed overview and an exhaustive list of references related to the conformal bootstrap programme.}. Thus, generally speaking, the bootstrap method in its current incarnation aims to extract numerical or analytical information, or general mathematical structures in theories that are not exactly solvable.

Despite the success of the conformal bootstrap method when applied to theories in AdS, extending the procedure to dS is challenging.  Complications are due to the fact that dS lacks a notion of spatial infinity and while the dS isometries act on the late-time boundary of dS space as $SO(d+1,1)$, the corresponding CFT is Euclidean that is typically non-unitary in a Lorentzian sense. The latter is a crucial difference with respect to the AdS counterpart, also because the existence of a convergent expansion of a bulk operator in terms of boundary operators is not guaranteed unlike in AdS/CFT. However in recent years, there have been several attempts to use symmetries, locality, and unitarity to constrain and characterise correlators on these theories. This progress is often referred to as the cosmological bootstrap. 

In this paper we follow an alternative route. Inspired by \cite{Gadde:2017sjg}, we write down a generalised group-theoretic crossing equation for a unitary (as opposed to reflection positive) Euclidean CFT and construct an exact solution to it. This crossing equation is of particular importance in the cosmological bootstrap programme \cite{Hogervorst:2021uvp, DiPietro:2021sjt} as it imposes constraints on the CFT living on the late time boundary of dS. These boundary conformal correlators in turn provide us with an approximation for the cosmological correlators evaluated in the far future. The associated quantum numbers are different from those in the standard crossing equation appearing in the conformal bootstrap programme, and are intricately related to harmonic analysis on the Euclidean conformal group $SO(d+1,1)$. In this paper, instead of trying to use crossing and unitarity to extract information on a physically relevant theory, we construct an exact solution to crossing that is manifestly unitary. The solution is based on the 6j symbols of the conformal group, and the crucial step here is the derivation of the analogue of the Biedenharn-Elliot identity or the pentagon identity for these 6j symbols. With appropriately chosen parameters, the pentagon identity can be morphed into the crossing equation, thereby giving us a solution where the dynamical three-point function coefficients of the theory are some specific 6j symbols. 

Thus we present here in closed form, a (potentially) interacting solution to the crossing equation in a unitary Euclidean CFT in $d \geq 2$ and to our knowledge it is the first example of its kind in $d>2$. On the flip side, the physical interpretation of this solution is not immediately clear and we discuss this briefly later in the paper. We study the analytic structure of the solution in $d=2$ which hints at a possible holographic interpretation.

%, but unfortunately we are unable to close the discussion definitively at this point.  

The structure of the paper is as follows. In Section \ref{sec:creq} we discuss the harmonic analysis of the Euclidean conformal group $SO(d+1,1)$, its irreducible representations and its relevance in the context of describing quantum field theories on dS spaces. In Section \ref{sec:soln} we provide a detailed derivation of the pentagon identity for the 6j symbols of the conformal group and the resulting solution to the crossing equation in terms of 6j symbols, and briefly discuss the analytic structure of the solution in two dimensions. Section \ref{sec:conc} contains concluding remarks and comments. Appendices \ref {sec:tsconv}, \ref{sec:6j2dexp} and \ref{sec:cpwcb} contain several technical details.

%%%%%%%%%%%%%%%%%%%%%%%%%%%%%%%%%%%%%%%%%%%%%%%%%%%%%%%%%%%%%%%%%%%%%%%%%%%%%%%%%%%%%%%%%%
\section{The crossing equation in conformal and cosmological bootstrap}
\label{sec:creq}
%%%%%%%%%%%%%%%%%%%%%%%%%%%%%%%%%%%%%%%%%%%%%%%%%%%%%%%%%%%%%%%%%%%%%%%%%%%%%%%%%%%%%%%%%%

We begin sec.\ref{sec:reps}  with a brief discussion on harmonic analysis of the Euclidean conformal group $SO(d+1,1)$, focusing on its unitary irreducible representations (UIR) . Later in sec. \ref{sec:cosmoboot}, we discuss how this is important in the context of quantum field theory in de Sitter spacetime (with the isometry group $SO(d+1,1)$).

%%%%%%%%%%%%%%%%%%%%%%%%%%%%%%%%%%%%%%%%%%%%%%%%%%%%%%%%%%%%%%%%%%%%%%%%%%%%%%%%%%%%%%%%%%
\subsection{Conformal representation theory}
\label{sec:reps}
%%%%%%%%%%%%%%%%%%%%%%%%%%%%%%%%%%%%%%%%%%%%%%%%%%%%%%%%%%%%%%%%%%%%%%%%%%%%%%%%%%%%%%%%%%

Harmonic analysis is a generalisation of Fourier analysis to topological groups, the main purpose being to decompose the regular representation of a group into irreducible ones. Since the regular representation is unitary, the irreps appearing in the decomposition are UIRs of the group. For a compact group $G$, every unitary representation is a direct sum of irreps and every irrep is finite dimensional. As stated by the Peter-Weyl theorem \cite{Weyl:1927}, the regular representation on $L^{2}(G)$ decomposes into a finite sum of all UIRs with each isomorphism class of UIRs appearing in the decomposition with multiplicity equal to the degree of the UIR. Additionally, the matrix coefficients of the UIRs (choosing one representative from each isomorphism class) provide an orthonormal basis for $L^{2}(G)$. The simplest example to consider would be $G=U(1)$. The UIRs are labelled by integer $n$, and we know all normalisable periodic functions can be decomposed in a Fourier series in terms of the matrix coefficients $e^{in\theta}$.

The representation theory analysis is more complicated for non-compact groups than it is for compact ones, and typically there is a continuous series of irreps involved. For the Euclidean conformal group in particular, there is a principal series of irreps $\Delta = \frac{d}{2} + i\nu$ with $\nu \in \mathbb{R}^{+}$ appearing in the decomposition of the regular representation\footnote{In $d=1$ there are in fact two different principal series of irreps \cite{Hogervorst:2021uvp}, however we shall only be interested in $d\geq 2$ in the present work.}, and additionally a complementary series, exceptional series, and discrete series of irreps the details of which depend on the conformal dimension and the spin. In analogy with the matrix coefficients in the compact case, the Conformal Partial Waves (CPWs) for the principal series irreps provide an orthonormal basis for decomposing four-point conformal correlators of scalar operators on the principal series.

For a connected semisimple real Lie group with a finite centre, the UIRs appearing in the decomposition of the regular representation are those that are induced from a UIR of the subgroup $P=MAN$\footnote{$P$ is called a parabolic subgroup, and $MAN$ is its Langlands decomposition.}, where $A$ and $N$ are the subgroups appearing in the Iwasawa decomposition $G=KAN$ and $M$ is the centraliser of $A$ in $K$ \cite{bams/1183531812, Gadde:2017sjg}. For $G=SO(d+1,1)$, $K$ is the maximal compact subgroup $SO(d+1)$, $A$ is the group of dilatations, $N$ is the group of special conformal transformations, and $M=SO(d)$. The inducing representation (say $\sigma$) of the subgroup $P$ transforms trivially under $N$ and is really an irrep of $MA$. Hence the UIRs of $SO(d+1,1)$ are labelled by the scaling dimension and the spin. The induced representation Ind$_{P}^{G}(\sigma)$ acts on the vector space $F$ of (compactly supported) sections of a homogeneous vector bundle on the base space $G/P$\footnote{As mentioned in \cite{Gadde:2017sjg}, $G/P$ for $SO(d+1,1)$ is isomorphic to the conformal compactification of $\mathbb{R}^{d}$ and that's how position space emerges from these group theoretic considerations.}. $\sigma$ being unitary allows us to define an inner product on $F$ completing it to a Hilbert space.

Harmonic analysis for the Euclidean conformal group was first studied in \cite{DOBREV1976219, Dobrev:1977qv}, with a recent resurgence of interest in the context of CFTs \cite{Karateev:2018oml}, and in studying QFT in dS spacetime \cite{Basile_2017,Sun:2021rrs,Sun:2021thf,Penedones:2023uqc}. We shall quote here the list of UIRs relevant in $d\geq 2$\footnote{We refer the reader to \cite{Penedones:2023uqc} for a more detailed discussion on the topic.} before moving on to focus on the special case of a four-point function of scalars: 
\begin{itemize}
\item \textbf{Principal series}:\\
	As mentioned before, these consist of the irreps with $\Delta = \frac{d}{2} + i\nu$ with $\nu\in \mathbb{R}^{+}$ for all spin $\ell$. The irreps $[\Delta,\ell]$ are isomorphic to $[d-\Delta,\ell]$ and therefore we could equivalently choose $\mathbb{R}^{-}$ (or $\mathbb{R}$) as the range of $\nu$.
\item \textbf{Complementary series}:\\
	These consist of the quantum numbers $\Delta \in \left(\frac{d}{2},d\right)$ for $\ell = 0$, and $\Delta \in \left(\frac{d}{2},d-1\right)$ for $\ell\geq 1$. Like in the principal series, the isomoprhy between $[\Delta,\ell]$ and $[d-\Delta,\ell]$ applies here as well. There are also certain analytic continuations of the complementary series called the \textbf{Exceptional series} that are relevant in the decomposition (being distinct from the other UIRs) for $d\geq 4$.
\item \textbf{Discrete series}:\\
	These consist of integer or half integer values of the scaling dimension, and only appear in odd dimensions. In $d=3$, these are the same as the Exceptional series \cite{Dobrev:1977qv,Sun:2021thf}. 
\end{itemize}   

We restrict our attention to correlation functions of scalar operators. For such irreps on the principal series, the only irreps appearing in the Clebsch-Gordan decomposition are the ones on the principal series\footnote{If we have a correlation function with pairwise identical scalars, we'd need to include the contribution of the unit operator in the appropriate channel(s). The connected part can still be decomposed in CPWs on the principal series.}\footnote{Also look at \cite{Mazac:2018qmi,Hogervorst:2021uvp} for the $d=1$ case where the discrete series is needed in addition to the principal series.} \cite{DOBREV1976219,Naimark1959,Martin1981,Penedones:2023uqc}. Thus we can use the following partition of unity,
\beq
\mathds{1} = \sum_{\ell} \int_{\frac{d}{2}}^{\frac{d}{2}+i\infty} \frac{d\Delta}{2\pi i}\; \mu(\Delta,\ell) \int_{\mathbb{R}^{d}} dx \ket{\Delta,\ell,x}_{\{\rho_{\ell}\}} \bra{\Delta,\ell,x}^{\{\rho_{\ell}\}}   		\quad  + \quad \text{discrete contributions},                           \label{unitypart}
\eeq
in $SO(d+1,1)$ correlation functions to expand them in the basis provided by the principal series. In the above $\ell$ specifies the $SO(d)$ irrep, and $\{\rho_{\ell}\}$ refers to the corresponding spin indices\footnote{We  will suppress these indices from now on. For integer spins, which are sufficient in our case, we shall assume that the spin indices are upper indices unless specified otherwise.}. $\ket{\Delta,x}$ is a state that transforms as one raised by a local primary acting on the vacuum $O_{\Delta,\ell}(x)\ket{\Omega}$\footnote{We refer the reader to \cite{Hogervorst:2021uvp} for a rigorous construction of these states.}. Similarly, $\bra{\Delta,x}$ transforms as $\left(O_{\Delta,\ell}(x)\ket{\Omega}\right)^{\dagger}$ which for bosonic operators is just $\bra{\Omega}O_{\Delta^{*},\ell}(x)$ with the spin indices lowered\footnote{The dagger maps to the dual spin representation. Since we shall be working with bosonic representations only, and will be keeping the indices implicit as well, from now onward we can just take the $\dag$ to mean complex conjugation of the scaling dimension.}. If $\Delta$ is on the principal series (as is the case here), then $O_{\Delta^{*},\ell}$ is the shadow $\tilde{O}_{\Delta,\ell}$ of $O_{\Delta,\ell}$ and has dimension $d-\Delta$. $\mu(\Delta,\ell)$ is the Plancherel measure \cite{Karateev:2018oml}, see App. \ref{sec:cpwcb} for more details. As mentioned before, we will be able to ignore any discrete contributions for the most part. 

Plugging \eqref{unitypart} in a four-point function of scalars, we obtain,
\bea
\braket{O_1 (x_1) O_2 (x_2) O_3 (x_3) O_4 (x_4)} = & \sum_{\ell_R } \int_{\frac{d}{2}}^{\frac{d}{2}+i\infty} \frac{d\Delta_R }{2\pi i} \; \mu(\Delta_R,\ell_R) \; K_{12R}K_{34\tilde{R}} \; \\
				& \int_{\mathbb{R}^{d}} dx \braket{O_1 (x_1) O_2 (x_2) O_R (x)} \braket{\tilde{O}_{R}(x) O_3 (x_3) O_4 (x_4)} .                   
\eea{cpwexp1}
In the above, we have introduced the three-point function coefficients,
\bea
\braket{O_1 (x_1) O_2 (x_2)| \Delta_R ,\ell_R ,x} & =  K_{12R} \braket{O_1 (x_1) O_2 (x_2) O_R (x)}    ,     \\
\therefore   \braket{\Delta_R ,\ell_R ,x | O_3 (x_3) O_4 (x_4) }  & =  K^{*}_{3^{\dag}4^{\dag}R} \braket{\tilde{O}_{R}(x) O_3 (x_3) O_4 (x_4)} = K_{34\tilde{R}} \braket{\tilde{O}_{R}(x) O_3 (x_3) O_4 (x_4)} .
\eea{3ptcoeff}
The position integral in \eqref{cpwexp1} is a CPW, and this gives us the CPW expansion of the four-point function on the principal series,
\beq
\braket{O_1 (x_1) O_2 (x_2) O_3 (x_3) O_4 (x_4)} =  \sum_{\ell_R } \int_{\frac{d}{2}}^{\frac{d}{2}+i\infty} \frac{d\Delta_R }{2\pi i} \; \mu(\Delta_R,\ell_R) \; K_{12R}K_{34\tilde{R}} \; \Psi^{R}_{1234}(x_i ) .             \label{cpwexp2}
\eeq
Since we are considering a correlator of scalars, the three-point function coefficients are non-zero only when $\ell_R$ is an integer spin. Thus there is also a unique tensor structure associated with the three -point function. The tensor indices on the three-point functions in \eqref{3ptcoeff} are mutually contracted and thus we get a scalar CPW $\Psi^{R}_{1234}(x_i )$. In Euclidean signature the CPW is a single valued function of position variables, and is an eigenfunction of the quadratic Casimir operator \cite{Dolan:2003hv}. CPWs on the principal series are orthogonal to each other,
\beq
\big( \Psi^{R}_{1234}, \Psi^{\tilde{R}}_{\tilde{1}\tilde{2}\tilde{3}\tilde{4}} \big)  = 2\pi \delta(\nu - \nu') \delta_{\ell_R \ell_{R'}} n_R  ,   \label{ortho}
\eeq
where $\Delta_R = \frac{d}{2}+i\nu$, and we have used the notation\footnote{We have to divide by $\text{Vol}SO(d+1,1)$ in order to compensate for integrating over redundant configurations (related by conformal transformations). Since the conformal group is non-compact, the integral would not be well-defined without this factor.},
\beq
\big( F, G \big) = \int \frac{dx_1 \cdots dx_n}{\text{Vol}SO(d+1,1)}  F(\{x_i\}) G(\{x_i\}) .      \label{pairing}
\eeq 
$n_R$ is a proportionality constant\footnote{We shall not need the explicit expression for $n_R$ but this can be found in \cite{Karateev:2018oml}, or in \cite{Simmons_Duffin_2018}, for example.}. 
 
 Note that \eqref{cpwexp2} is an $s$-channel expansion as we inserted the complete basis between $\bra{O_1 O_2}$ and $\ket{O_3 O_4}$. We can do the same in other channels to obtain the corresponding CPW expansions. The completeness of the basis implies that these different expansions should be equal to each other. This gives us a set of constraints on the unknown three-point function coefficients. For example, the equality of the $s$-channel and the $t$-channel expansions gives us the following crossing equation \cite{Hogervorst:2021uvp}\,
\beq
\sum_{\ell_R } \int_{\frac{d}{2}}^{\frac{d}{2}+i\infty} \frac{d\Delta_R }{2\pi i} \mu(\Delta_R,\ell_R) K_{12R}K_{34\tilde{R}} \; \Psi^{R}_{1234}   =    \sum_{\ell_{R'} } \int_{\frac{d}{2}}^{\frac{d}{2}+i\infty} \frac{d\Delta_{R'} }{2\pi i} \mu(\Delta_{R'},\ell_{R'}) K_{32R'}K_{14\tilde{R}'}  \; \Psi^{R'}_{3214}     .      \label{stcrossing}
\eeq

As shown in \cite{Hogervorst:2021uvp}, the fact that $R$ is on the principal series imposes positivity constraints on the product of three-function coefficients $K_{12R}K_{34\tilde{R}}$ (that we call the spectral function $I_{1234}^{R}$ with the indices in subscript indicating the channel just as in the case of CPWs). When $O_1 = O_{3}^{\dag}$ and $O_2 = O_{4}^{\dag}$, we have the following property following from the unitarity of $R$,
\beq
K_{12R}K_{34\tilde{R}} = K_{12R}K_{1^{\dag}2^{\dag}\tilde{R}} = K_{12R}K_{1^{\dag}2^{\dag}R^{\dag}} = K_{12R}K^{*}_{12R}  \geq 0   .     \label{unitarity1} 
\eeq
Thus we have the (position dependent) crossing equation in \eqref{stcrossing} and a unitarity constraint in \eqref{unitarity1}. 

A CPW in one channel can be expanded in terms of CPWs in another channel with the expansion coefficient being a 6j symbol,
\beq
\Psi_{3214}^{R'} = \sum_J \int_{\frac{d}{2}}^{\frac{d}{2}+i\infty} \frac{d\Delta}{2\pi i} \frac{1}{n_R} \; \left\{ \begin{array}{ccc} 1 & 2 & R' \\  3 & 4 & R \end{array} \right\} \; \Psi_{1234}^{R}   . \label{cpwexp}
\eeq
The 6j symbol is defined as a conformal invariant pairing \eqref{pairing} of the CPWs in two different channels as follows, 
\beq
\left\{ \begin{array}{ccc} 1 & 2 & R' \\  3 & 4 & R \end{array} \right\}  = \Big(\Psi_{\tilde{1}\tilde{2}\tilde{3}\tilde{4}}^{\tilde{R}},  \Psi_{3214}^{R'}   \Big)    ,  \label{cpwpair} 
\eeq
or equivalently as the following conformally invariant integral of the product of 3-pt functions\footnote{Note that we are following the definitions in \cite{Liu:2018jhs}.},
\beq
\left\{ \begin{array}{ccc} 1 & 2 & R' \\  3 & 4 & R \end{array} \right\}^{(abcd)} = \int \frac{dx_1 \cdots dx_4 dx_R dx_{R'}}{\text{Vol}SO(d+1,1)}  \braket{\tilde{O}_1 \tilde{O}_2 \tilde{O}_R}^{(a)} \braket{O_R \tilde{O}_3 \tilde{O}_4}^{(b)} \braket{O_3 O_2 O_{R'}}^{(c)} \braket{\tilde{O}_{R'}O_1 O_4}^{(d)}    .   \label{6jdef}
\eeq
The three-point structures are not always unique, and hence the 6j symbol can generically carry labels for tensor structures. Since we have assumed the external operators to be scalars and the three-point structures between two scalars and a bosonic operator are unique, we do not need these labels. 
 
\begin{figure}
\centering
		\begin{tikzpicture}[anchor=base,baseline,scale=1.2]
			\node [coordinate] (145) at (0,0) {};
			\node [coordinate] (126) at (-3.3,-1) {};
			\node [coordinate] (364) at ( -1.5,-3) {};
			\node [coordinate] (235) at ( -2, -1.7) {};
			\node at (-1.8,-0.4) {$1$};
			\node at (-2.65, -2.2) [] {$6$};
			\node at (-0.5,-1.5) [] {$4$};
			\node at (-2.35,-1.4) [] {$2$};
			\node at (-1.7,-2.15) [] {$3$};
			\node at (-1.3,-0.9) [] {$5$};
			\draw [thick] (126) -- (145);
			\draw [thick] (145) -- (364);
			\draw [thick] (364) -- (126);
			\draw [thick, dashed] (364) -- (235);
			\draw [thick, dashed] (126) -- (235);
			\draw [thick, dashed] (145) -- (235);
		\end{tikzpicture} 
	\caption{Diagrammatic representation of the 6j symbol $\left\{ \begin{array}{ccc} 1 & 2 & 6 \\  3 & 4 & 5 \end{array} \right\}$ as a tetrahedron. }      \label{6jtetra}
\end{figure}
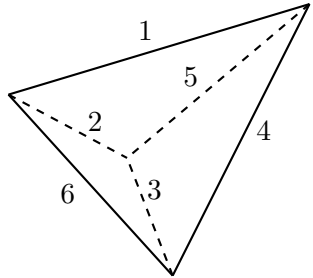

Let us now mention an important property of the 6j symbols that will be useful later. As it is evident from the definition \eqref{6jdef}, the 6j symbols enjoy, modulo some shadow transforms, a tetrahedral symmetry generated by the following,
\beq
\left\{ \begin{array}{ccc} 1 & 2 & 6 \\  3 & 4 & 5 \end{array} \right\} = \left\{ \begin{array}{ccc} \tilde{5} & 3 & 2 \\  6 & \tilde{1} & 4 \end{array} \right\} = \left\{ \begin{array}{ccc} 2 & 5 & \tilde{3} \\  \tilde{4} & 6 & 1 \end{array} \right\} = \left\{ \begin{array}{ccc} 2 & 1 & \tilde{6} \\  4 & 3 & 5 \end{array} \right\}       ,   \label{tetra}
\eeq
and can be diagrammatically represented as the tetrahedron in fig. \ref{6jtetra}\footnote{Note that the tetrahedron in fig. \ref{6jtetra} is the dual-tetrahedron in the conventions followed in Fig. 6 in \cite{Liu:2018jhs}.}.

We can plug \eqref{cpwexp} in to the r.h.s. of \eqref{stcrossing}, and then use the orthogonality of the CPWs in \eqref{ortho} to eliminate the CPWs (in the same channel) from both sides. This gives us,
\beq
n_{R} K_{12R}K_{34\tilde{R}}  = \sum_{J'} \int_{\frac{d}{2}}^{\frac{d}{2}+i\infty} \frac{d\Delta'}{2\pi i}  K_{32R'}K_{14\tilde{R'}}  \; \left\{ \begin{array}{ccc} 1 & 2 & R' \\  3 & 4 & R \end{array} \right\}  \quad . \label{posind}
\eeq
This is the position independent crossing equation \cite{Gadde:2017sjg, Hogervorst:2021uvp} that is the central object of interest in this paper. In sec. \ref{sec:soln} we present an exact solution to this equation following in the footsteps of \cite{Gadde:2017sjg}.

So far we have only discussed the UIRs of the Euclidean conformal group $SO(d+1,1)$ and in particular the principal series. However these are only of indirect interest in Lorentzian CFTs where we are concerned about the UIRs of the Lorentzian conformal group that appear in the conformal block expansion of correlators. As we shall discuss in App. \ref{sec:cpwcb}, we can go from the CPW expansion in \eqref{cpwexp2} to the conformal block expansion, and the poles in the spectral function $I_{1234}^{R}$ capture the physical spectrum of a theory.

However, \eqref{stcrossing} does impose constraints on the conformal theory living on the late time boundary of de Sitter spacetime and it is thus relevant in cosmological bootstrap \cite{Hogervorst:2021uvp}. This is what we shall briefly review in the next section. Experts on the topic can safely skip the section to go on to sec. \ref{sec:soln}.

%%%%%%%%%%%%%%%%%%%%%%%%%%%%%%%%%%%%%%%%%%%%%%%%%%%%%%%%%%%%%%%%%%%%%%%%%%%%%%%%%%%%%%%%%%
\subsection{Cosmological bootstrap}
\label{sec:cosmoboot}
%%%%%%%%%%%%%%%%%%%%%%%%%%%%%%%%%%%%%%%%%%%%%%%%%%%%%%%%%%%%%%%%%%%%%%%%%%%%%%%%%%%%%%%%%%

In this section, we briefly review how \eqref{posind} is relevant in the context of ``cosmological bootstrap" \cite{Hogervorst:2021uvp,DiPietro:2021sjt,Baumann:2022jpr}. 

\begin{figure}
	\begin{center}
	\begin{tikzpicture}[anchor=base,baseline]
		\node (yb) at (0,-2.5) [] {};
		\node (yt) at (0, 2.5) [] {};
		\node (xl) at ( -2.5, 0) [] {};
		\node (xr) at ( 2.5,0) [] {};
		\node (tl) at (-2,2) [] {};
		\node (br) at (2,-2) [] {};
		\draw [-stealth] (yb) -- (yt);
		\draw [-stealth] (xl)-- (xr);
		\draw [dashed] (br) -- (tl);
		\node at (3,0) {$X^{d+1}$};
		\node at (0,2.6) {$X^{0}$};
		\node at (1.5,1.2) {$X^{0} + X^{d+1} \geq 0$};
		\node at (-1.5,-1.2) {$X^{0} + X^{d+1} < 0$};
		\pgfmathsetmacro{\e}{1.5}   % eccentricity
    		\pgfmathsetmacro{\a}{1}
    		\pgfmathsetmacro{\b}{(\a*sqrt((\e)^2-1)} 
    		\draw[thick] plot[domain=-1.5:1.5] ({\a*cosh(\x)},{\b*sinh(\x)});
    		\draw[thick] plot[domain=-1.5:1.5] ({-\a*cosh(\x)},{\b*sinh(\x)});
	\end{tikzpicture}
	\end{center}
	\caption{The hyperbola is a two-dimensional projection of dS$_{d+1}$ hypersurface. The lightlike dotted line separates the two causally indepedent patches $X^{0} + X^{d+1} \geq 0$ and $X^{0} + X^{d+1} < 0$. We'll focus on the former and use the Poincare coordinates defined in \eqref{Poincare} for this patch.}
	\label{fig:dSpatches}
	\end{figure}
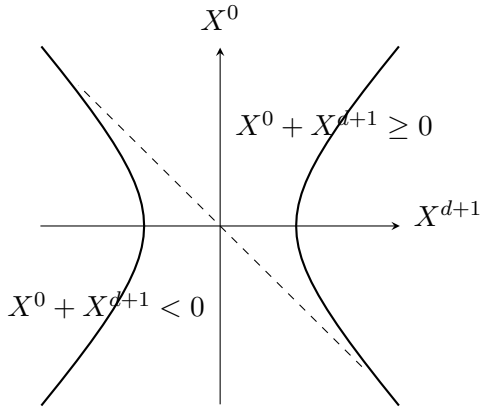

Empty de Sitter spacetime dS$_{d+1}$ is a maximally symmetric spacetime and can be described as the following hypersurface embedded in $d+2$ dimensional Minkowski spacetime $\mathbb{M}^{d+1,1}$, 
\beq
-(X^0 )^2 + (X^1 )^2 + \cdots + (X^{d+1} )^2 = R^2    ,       \label{dSembed}
\eeq
where $R$ is the Hubble radius. We can patch all of dS with the following global coordinates,
\beq
X^0 = R \sinh t , \quad \quad X^i = R \; y^i \cosh t  \quad \forall\; i \in \{1, \cdots, d+1\} .     \label{dSglobal}
\eeq
We however focus on the Poincare coordinates that patch only half of global dS. As shown in fig. \ref{fig:dSpatches}, we can split global dS into two causally independent patches $X^0 + X^{d+1} \geq 0$ and $X^0 + X^{d+1}<0$.
We can use the following Poincare coordinates $\{\eta,x^{\mu}\}$ for $X^0 + X^{d+1} \geq 0$,
\beq
X^0 = \frac{R}{\eta} \left(\eta^2 -1 - x^2\right) , \quad X^{d+1} = \frac{R}{\eta} \left(x^2 -1 - \eta^2 \right), \quad X^{\mu} = -\frac{R}{\eta} x^{\mu}, \quad \eta<0, \;\; x^{\mu} \in \mathbb{R}^{d} .       \label{Poincare}
\eeq
The metric on this patch in these coordinates is as follows,
\beq
ds^2 = R^2 \frac{d\eta^2 - dx^2 }{\eta^2} .         \label{Poincmetric}
\eeq
The late time boundary $\eta \rightarrow 0$ (and any constant $\eta$ slice) is conformally equivalent to $\mathbb{R}^d$.

The dS isometry group is $SO(d+1,1)$. Hence, states on the Hilbert space corresponding to a dS field $\phi(\eta,x)$ can be labelled with the dilatation eigenvalue $\Delta_{\phi}$ and $SO(d)$ spin $\ell_{\phi}$. For a scalar particle, the mass is related to $\Delta$ as follows,
\beq
m^2 R^2 = \Delta_{\phi}(d-\Delta_{\phi})        . \label{massdict}
\eeq
In a free unitary theory, $m^2 \geq 0$, and we can now see that $\Delta_{\phi}$ is in the complementary series when $0 \leq m^2 R^2 \leq \frac{d^2}{4}$ (light fields) and in the principal series when $m^2 R^2 \geq \frac{d^2}{4}$ (heavy fields).

The objects of interest in studying QFT in dS are correlation functions of local operators defined w.r.t. the Bunch-Davies vacuum \cite{1968AnHP,Bunch1978}. The feature that is of utmost important to us is that these correlators can be calculated in an expansion consisting of correlators defined on the late time boundary $\eta\rightarrow 0$. This is basically exploiting the following expansion of bulk fields $\phi(\eta,x)$ in terms of boundary operators $O(x)$,
\beq
\phi(\eta,x) = \sum_{O} b_{\phi O} (-\eta)^{\Delta} \left[O(x) + c_1 \eta^{2} \partial_{x}^{2} O(x) + \cdots \right] ,    \label{bulkbdy}
\eeq
where $b_{\phi O}$ are dynamically determined coefficients but the coefficients $c_i$ are fixed by the $SO(d+1,1)$ isometry. One can check the commutation relations of the $SO(d+1,1)$ generators with the boundary operators and this tells us that $O(x)$ transforms as a primary operator in a CFT living on the late time boundary with dimension $\Delta$\footnote{Note that $\Delta$ is different from $\Delta_{\phi}$ and is not limited to the UIRs discussed previously and can be any complex number.} (the following operators in the same series being descendants)\footnote{The existence of the expansion in \eqref{bulkbdy} is not on solid ground as in AdS/CFT. Roughly speaking this is because a hypersurface anchored on the late time boundary that respects the dS isometries is non-compact and hence there is no state-operator correspondence going from a state in the dS QFT Hilbert space to an operator in the CFT on the late time boundary. Note that in the case of AdS/CFT, a Hilbert space defined on a constant $\tau$ slice of global AdS can be mapped to a compact hemisphere on the boundary of AdS using the same isometries, and then the scale symmetry can be used to argue that this defines a local operator in the boundary CFT. We refer the reader to \cite{DiPietro:2021sjt} for a detailed argument, and to sec. 2.3 of \cite{Hogervorst:2021uvp} for an explicit example to demonstrate the same point.}. 

This boundary CFT is the objective of this paper. The CFT in question is not a standard physically interesting CFT as the dimensions of the operators are not subject to the unitarity constraints in \eqref{unitaritybounds}. All this while we have also assumed that the gravity in dS is not dynamical and hence the boundary CFT does not have a conserved stress energy tensor. Correlators in the theory can be however expanded in terms of CPWs on the principal series just as in \eqref{cpwexp2}. As shown in App. \ref{sec:cpwcb}, the CPW expansion also gives a conformal block expansion although in this case this expansion cannot be interpreted in terms of a convergent OPE\footnote{We refer the reader to sec. 4.3 of \cite{DiPietro:2021sjt} for an interpretation in terms of quasi-normal modes in dS.}.

We can subject the correlators in this theory to the bootstrap constraint in \eqref{posind} such that it satisfies \eqref{unitarity1}. Any results obtained therefrom, by virtue of \eqref{bulkbdy}, amounts to non-perturbative results for the cosmological correlators of the QFT in the bulk dS spacetime. In the next section, we construct one exact solution to the position independent crossing equation \eqref{posind}.

 %As previously mentioned, we shall however not endeavour to use the crossing constraints to put bounds on CFT data or to devise approximation schemes for generic theories. 

%%%%%%%%%%%%%%%%%%%%%%%%%%%%%%%%%%%%%%%%%%%%%%%%%%%%%%%%%%%%%%%%%%%%%%%%%%%%%%%%%%%%%%%%%%
\section{Constructing a solution to the crossing equation}
\label{sec:soln}
%%%%%%%%%%%%%%%%%%%%%%%%%%%%%%%%%%%%%%%%%%%%%%%%%%%%%%%%%%%%%%%%%%%%%%%%%%%%%%%%%%%%%%%%%%

We would like to start by introducing 6j symbols as recoupling coefficients in tensor products of conformal irreps. To this end, let us take the scalar operators $O_1$ and $O_2$, and consider the tensor product of the irreps of the corresponding states. In a short while we generalise this (partially) to include spins. We shall use the partition of unity in \eqref{unitypart} to expand the tensor product of these states as follows\footnote{As explained once previously, $\ket{\Delta,\ell,x} = O(x)\ket{\Omega}$ and $\bra{\Delta,\ell,x} = \bra{\Omega} \tilde{O}(x)$ (when $O$ is on the principal series).}\footnote{Note that we have shortened the notation for the Plancherel measure.},
\bea
\ket{O_1(x_1)}\otimes \ket{O_2(x_2)}  = & \sum_{\ell} \int_{\frac{d}{2}}^{\frac{d}{2}+i\infty} \frac{d\Delta}{2\pi i}\; \mu_O \int dx \ket{O(x)} \bra{\tilde{O}(x)} \left(\ket{O_1(x_1)} \otimes \ket{O_2(x_2)}\right)  ,   \\
		=& \sum_{\ell} \int_{\frac{d}{2}}^{\frac{d}{2}+i\infty} \frac{d\Delta}{2\pi i}\; \mu_O \int dx \ket{O(x)} \braket{\tilde{O}(x)O_1(x_1)O_2(x_2)} ,
\eea{CG}
\eqref{CG} is true\footnote{Note that the product $\otimes$ between $\ket{O_1(x_1)}$ and $\ket{O_2(x_2)}$ is a tensor product of representations and not a Hilbert space tensor product. We thank A. Manenti for pointing this out to us. In the same vein, $\bra{\tilde{O}(x)} \left(\ket{O_1(x_1)} \otimes \ket{O_2(x_2)}\right)$ is not a Hilbert space inner product and hence it is not a 3-pt correlation function but simple a tensor structure. We expand on how to define/count tensor structures in representation theory terms shortly.} when $O_1$ and $O_2$ are on the principal series themselves, but we shall assume that we can make an analytic continuation of the final results to generic complex values of dimensions for $O_1$ and $O_2$\footnote{We have dropped the discrete terms from \eqref{unitypart} in \eqref{CG} because (as explained before) these will not play a role in our analysis other than for specific correlators with a disconnected part.}.

Note that $\braket{O(x)O_1(x_1)O_2(x_2)}$ in the second line does not refer to a three-point function with a dynamical coefficient. \eqref{CG} is a statement about group representations and $\braket{O(x)O_1(x_1)O_2(x_2)}$ is merely a three-point tensor structure\footnote{We shall use $\braket{\cdots}$ to refer to tensor structures unless otherwise mentioned.}. \eqref{CG} can be thought of as the analogue of Clebsch-Gordon decomposition in the case of the Euclidean conformal group. 

Since $O_1$ and $O_2$ are scalars, $O$ is restricted to being an integer spin operator, and there is a unique tensor structure \cite{Costa:2011mg}. $[\Delta,\ell]$ are the quantum numbers of $O$. 

We shall now use the following bubble integral \cite{Karateev:2018oml} to invert \eqref{CG} so we can express the state $\ket{O(x)}$ in terms of the tensor product on the l.h.s.
\beq
\int dx_1 \int dx_2 \braket{O_R(x) O_1 (x_1 ) O_2 (x_2 ) } \braket{\tilde{O}_{R'}(x') \tilde{O}_1 (x_1 ) \tilde{O}_2 (x_2 ) } = 2\pi \;m_{R,R} \; \delta(\nu_R - \nu_{R'}) \delta_{\ell_R, \ell_{R'}} \mathbf{1}_{xx'}   .  \label{bubble}
\eeq  
Quantum numbers of $O_R$ are $[\frac{d}{2}+ i\nu_R, \ell_R]$ and similarly for $O_{R'}$. The result above is entirely a consequence of the irreducibility and inequivalence of the principal series representations on which $O_R$ and $O_{R'}$ lie\footnote{We refer the reader to sec. 2.6 of \cite{Karateev:2018oml} for a discussion on bubble diagrams.}. The labels in $m_{l,ijk}$ are as per the following convention: $O_l$ is the leg not being contracted in \eqref{bubble} while $i$,$j$,$k$ refer to the legs with spin. When any one of the legs, say $k$, is a scalar, we simply drop the corresponding label and write it as $m_{l,ij}$. In \eqref{bubble}, the only operator with spin is $O_R$. $m_{R,R}$ can be obtained by tracing over the free legs in the bubble integral \eqref{bubble}\footnote{We refer the reader to sec. 2.6 of \cite{Karateev:2018oml} for the detailed derivation of $m_{R,R}$ (note the difference in notation).} and is independent of any contracted scalar legs. It is evident that $m_{l,ijk}$ is invariant under taking any of the operators to the corresponding shadow.  Inverting \eqref{CG} using the bubble integral gives us,
\beq
\ket{O(x)} = \frac{1}{C_{O}} \int dx_1 \int dx_2 \braket{O(x) \tilde{O}_1 (x_1 )\tilde{O}_2 (x_2 ) } \; \ket{O_1(x_1)}\otimes \ket{O_2(x_2)}     ,       \label{rCG}
\eeq
where $C_{O} = m_{O,O} \mu_O$. As we shall see later in \eqref{3pairab}, $C_O$ is a pairing of three-point structures. In the general spinning case that we will talk about shortly, we follow the notation $C_{ijk} = m_{O,ijk} \mu_O$ since the labels on $C_{ijk}$ refer to the spinning legs involved, and just as for $m_{l,ijk}$ we drop the labels for any of the legs that is a scalar. The $SO(d)$ indices of $O(x)$ on the l.h.s. of \eqref{rCG} are carried by the three-point structure on the r.h.s. 

So far we have assumed that the operators $O_1$ and $O_2$ are scalars. Let us now assume that one of them, say $O_1$, has integer spin\footnote{We can consider a more general situation where either of $O_1$ or $O_2$ could be in a generic irrep of $SO(d)$ but we are going to stick to the simpler case of $O_1$ having integer spin and $O_2$ being a scalar since this allows us to illustrate the main principles involved in the spinning generalisation while avoiding clutter.}, and let us go over the steps \eqref{CG} through \eqref{rCG} and the modifications necessary to suit this case.

Consider the second line in \eqref{CG}. As mentioned previously $\braket{\tilde{O} O_1 O_2}$ is a three-point structure. When $O_1$ has integer spin, the operator $O$ is also allowed to have integer spin such that the three-point function is non-zero and the three-point structure is no longer unique. The space of tensor structures has been studied in detail in \cite{Kravchuk:2016qvl}. The central idea is that the n-point tensor structures should be invariant under the little group (stabiliser) for the n-point function, and hence they are in one-to-one correspondence with the singlets of the little group appearing in the tensor product of the $SO(d)$ irreps in the correlator. If we are considering a correlator of operators $O_i$ in $SO(d)$ irreps $\rho_i$, then the space of tensor structures is,
\beq
\left(\text{ Res}^{\;SO(d)}_{\text{Little group}} \;\;\; \tens{i} \; \rho_i \right)^{\text{Little group}}       ,          \label{structures}
\eeq 
where $\text{Res}^{G}_{H}$ refers to the restriction of a representation of the group $G$ in terms of irreps of the subgroup $H$, and $(\tensadd{i} \;\; \rho_i)^{H}$ refers to the selection of $\rho_i$ that are singlets under the action of $H$. The stabiliser of the three-point function is $SO(d-1)$, therefore the space of three-point structures is given by,
\beq
\left(\text{ Res}^{\;SO(d)}_{SO(d-1)} \;\;\; \rho_1 \otimes \rho_2 \otimes \rho_3 \right)^{SO(d-1)}       .          \label{structures3pt}
\eeq 
Having multiple structures means we now have an index for the three-point structures in \eqref{CG} that ranges over this space in \eqref{structures3pt}. 

Since the three-point structures have a natural interpretation in terms of singlets of $SO(d-1)$, we shall also look at both sides of \eqref{CG} after taking a restriction to $SO(d-1)$. To illustrate this, let us consider the decomposition of the $SO(d)$ irrep $O_1$ is in, say $\rho_1$, in terms of irreps of $SO(d-1)$,
\beq
\text{Res}^{SO(d)}_{SO(d-1)} \; \rho_1 = \tensadd{k} \; \sigma_{1,k}     \quad .       \label{restric} 
\eeq
Let us pick one particular $\sigma_{1,k}$ from the decomposition in \eqref{restric}. In order to form a $SO(d)$ singlet, we must therefore pick $\sigma_{1,k}^{*}$ from the corresponding decomposition for $\bra{\tilde{O}}$\footnote{Assuming $\sigma_{1,k}$ does appear in the $SO(d-1)$ restriction of $O$ as well, if not we'd just need to pick a different irrep from the restriction in \eqref{restric}.} where the $ ^{*}$ indicates the dual representation. $O_2$, as per our assumptions, is a scalar. We can thus label the tensor structures with $k$. With these choices, we can now write the spinning version of \eqref{CG} as follows,
\bea
\ket{O_1(x_1)}_{\sigma_{1,k}}\otimes \ket{O_2(x_2)}  =  \sum_{\ell_k} \int_{\frac{d}{2}}^{\frac{d}{2}+i\infty} \frac{d\Delta}{2\pi i}\; \mu_O \int dx \ket{O(x)}_{\sigma_{1,k}} \braket{\tilde{O}(x)O_1(x_1)O_2(x_2)}^{(k)} ,
\eea{CGspin}
where $\ell_k$ refers to the values of $\ell$ as the spin of $O$ that can provide the irrep $\sigma_{1,k}$ upon taking the restriction to $SO(d-1)$. $\ket{s}_\sigma$ refers to the $\sigma$ component of $\ket{s}$ after decomposing it into irreps of $SO(d-1)$. We can now sum both sides of \eqref{CGspin} over the different irreps on the r.h.s. of \eqref{restric}\footnote{This decomposition is free from multiplicities \cite{vilenkin2010representation}.} to obtain the generalisation of \eqref{CG} to the case when $O_1$ has integer spin.

A simple example portraying these principles is the case of three dimensions. The spin $j$ $SO(3)$ irrep decomposes into $SO(2)$ irreps with quantum numbers $q\in\{j,j-1,\cdots, -j\}$. For a generic spinning three-point structure we must have $q_1 +q_2 + q_3 = 0$ for invariance under the little group $SO(2)$. The three point structure is thus generically specified by two $q$ values $[q_1 , q_2]$. When we assume $O_2$ to be a scalar, we can simply use $q_1$. In this case, we can therefore write \eqref{CG}, using the familiar angular momentum notation from Quantum Mechanics, as follows,
\beq
\ket{\Delta_1, j_1 ,x_1}\otimes \ket{\Delta_2, j_2=0, x_2}  =  \sum_{q_1 = -j_1}^{j_1} \;\; \sum_{j \geq |q_1|} \; \int_{\frac{d}{2}}^{\frac{d}{2}+i\infty} \frac{d\Delta}{2\pi i}\; \mu_O \int dx \ket{\Delta, j, q_1 , x} \; \braket{\text{T.S.}}^{(q_1)}             ,      \label{threedspin}
\eeq  
where $\braket{\text{T.S.}}^{(q_1)}$ denotes the three-point structure labelled by $q_1$.

In the next step, we need to generalise the bubble integral in \eqref{bubble}. In the case of spinning correlators with more than one three-point structures, the bubble integral is defined w.r.t. the individual structures and the r.h.s. would be a matrix labelled by the tensor structure indices\footnote{In \eqref{bubblespin} and \eqref{bubblespin2}, we have abbreviated $\int dx_1 \int dx_2$ to $\int dx_{1,2}$.} as follows,
\beq
\int dx_{1,2} \braket{O_R(x) O_1 (x_1 ) O_2 (x_2 ) }^{(a)} \braket{\tilde{O}_{R'}(x') \tilde{O}_1 (x_1 ) \tilde{O}_2 (x_2 ) }^{(b)} = 2\pi \;[m_{R,R12}]_{ab} \; \delta(\nu_R - \nu_{R'}) \delta_{\ell_R, \ell_{R'}} \mathbf{1}_{xx'}   .  \label{bubblespin}
\eeq 

The matrix element $[m_{R,R12}]_{ab}$ is proportional to the three-point pairing (defined in \eqref{pairing}) corresponding to the bubble integral \cite{Karateev:2018oml},
\beq
\left(\braket{O_R(x) O_1 (x_1 ) O_2 (x_2 ) }^{(a)} \braket{\tilde{O}_{R}(x') \tilde{O}_1 (x_1 ) \tilde{O}_2 (x_2 ) }^{(b)}\right) =  \mu_{R} \; [m_{R,R12}]_{ab} = [C_{R12}]_{ab}   .  \label{3pairab}
\eeq
This matrix of three-point pairings, being a positive definite Hermitian inner product on the space of singlets in \eqref{structures3pt}, is non-degenerate and can be diagonalised\footnote{We refer the reader to sec. 2.3 of \cite{Karateev:2018oml} for a detailed discussion.}. Therefore we have,
\bea
\int dx_{1,2} \braket{O_R(x) O_1 (x_1 ) O_2 (x_2 ) }^{(a)} \braket{\tilde{O}_{R'}(x') \tilde{O}_1 (x_1 ) \tilde{O}_2 (x_2 ) }^{(b)} = & \quad 2\pi \;m_{R,R12}^{(a)}\delta_{ab^*} \\
	& \delta(\nu_R - \nu_{R'}) \delta_{\ell_R, \ell_{R'}} \mathbf{1}_{xx'}   . 
\eea{bubblespin2}
Henceforth, we shall use the notation $\mu_{R}m_{R,R12}^{(a)} = C_{R12}^{(a)}$. Note that $C_{ijk}$ (and consequently, $m_{l,ijk}$) are invariant under permutations of $\{i,j,k\}$.

As an example, let us once again refer to the three dimensional case for a simple illustration. With $O_2$ assumed to be a scalar, the three-point structures can be labelled by the quantum number $q_1$ as we discussed previously. $[m_{R,1R}]_{q_1 q}$ is non-zero only if $q = q_1^{*} = -q_1$. 

\eqref{CGspin} can be simply inverted using \eqref{bubblespin2} to obtain the analogue of \eqref{rCG}. This can be expressed similarly as in \eqref{CGspin},
\beq
\ket{O(x)}_{\sigma_{1,j}} = \frac{1}{C_{1O}^{(j)}} \int dx_1 \int dx_2 \braket{O(x) \tilde{O}_1 (x_1 )\tilde{O}_2 (x_2 ) }^{(j)} \; \ket{O_1(x_1)}_{\sigma_{1,j}}\otimes \ket{O_2(x_2)}     .      \label{rCGspin} 
\eeq

Note that we can also just take the tensor product of states $\ket{O_1}\otimes \ket{O_2}$ on the r.h.s. of \eqref{rCGspin} and obtain the same result on the l.h.s. as the diagonal nature of the bubble integral in the space of three-point structures ensures that the correct component of $\ket{O_1}$ in the decomposition into $SO(d-1)$ irreps will be picked up by the three-point structure in the integral. If we should choose the operator $O(x)$ to be a scalar by integrating both sides of \eqref{CGspin} against the appropriate three-point structure, we can also get away with having a unique three-point structure on the r.h.s. of \eqref{rCGspin} and we can drop the labels on the three-point structures for the $SO(d-1)$ irreps. 

Armed with \eqref{rCG} and \eqref{rCGspin}, we want to now look at the corresponding statement for a tensor product of three irreps and how we can extract a specific irrep from the decomposition into irreps of $SO(d)$. We'll consider a few different cases separately.\\

\begin{figure}[h]
	\begin{subfigure}[b]{\linewidth}
		\begin{equation*}
		\begin{tikzpicture}[anchor=base,baseline,scale=0.8]
			\node [coordinate] (BL) at (-1.5,-1) {};
			\node [coordinate] (BM) at (0, -1) {};
			\node [coordinate] (BR) at ( 1.5,-1) {};
			\node [coordinate] (M1) at ( -0.75, 0) {};
			\node [coordinate] (M2) at ( 0, 1) {};
			\node [coordinate] (T) at ( 0.75, 2) {};
			\node at (-1.5,-1.4) {$1$};
			\node at (0, -1.4) [] {$2$};
			\node at ( 1.5,-1.4) [] {$3$};
			\node at ( -0.6, 0.6) [] {$r$};
			\node at ( 1.05, 2.2) [] {$\ket{O}^{(r)}$};
			\draw [thick] (BL) -- (M1);
			\draw [thick,decorate, decoration=snake] (M1) -- (M2);
			\draw [thick] (M2) -- (T);
			\draw [thick] (BM) -- (M1);
			\draw [thick,decorate, decoration=snake] (BR) -- (M2);
		\end{tikzpicture} \,
		\quad   \sim \sum_{\ell_{r'}} \int_{\frac{d}{2}}^{\frac{d}{2}+i\nu} \frac{d\Delta_{r'}}{2\pi i} \mu_{r'} W_{1a} (O_1 O_2 O_3 \tilde{O}; O_r O_{r'})           \quad
		\begin{tikzpicture}[anchor=base,baseline,scale=0.8]
			\node [coordinate] (BL) at (-1.5,-1) {};
			\node [coordinate] (BM) at (0, -1) {};
			\node [coordinate] (BR) at ( 1.5,-1) {};
			\node [coordinate] (M1) at ( 0.75, 0) {};
			\node [coordinate] (M2) at ( 0, 1) {};
			\node [coordinate] (T) at ( 0.75, 2) {};
			\node at (-1.5,-1.4) {$1$};
			\node at (0, -1.4) [] {$2$};
			\node at ( 1.5,-1.4) [] {$3$};
			\node at ( 0.6, 0.6) [] {$r'$};
			\node at ( 1.05, 2.2) [] {$\ket{O}^{(r')}$};
			\draw [thick] (BL) -- (T);
			\draw [thick] (BM) -- (M1);
			\draw [thick,decorate, decoration=snake] (BR) -- (M1);
			\draw [thick, decorate, decoration=snake] (M2) -- (M1);
		\end{tikzpicture} \,
		\end{equation*}
	\caption{$O_1$, $O_2$, and $O$ are scalars, while the intermediates $O_r$ and $O_{r'}$ alongwith the external operator $O_3$ have integer spin.}      \label{recoup1}
	\end{subfigure}\\
	\begin{subfigure}[b]{\linewidth}
		\begin{equation*}
		\begin{tikzpicture}[anchor=base,baseline,scale=0.8]
			\node [coordinate] (BL) at (-1.5,-1) {};
			\node [coordinate] (BM) at (0, -1) {};
			\node [coordinate] (BR) at ( 1.5,-1) {};
			\node [coordinate] (M1) at ( -0.75, 0) {};
			\node [coordinate] (M2) at ( 0, 1) {};
			\node [coordinate] (T) at ( 0.75, 2) {};
			\node at (-1.5,-1.4) {$1$};
			\node at (0, -1.4) [] {$2$};
			\node at ( 1.5,-1.4) [] {$3$};
			\node at ( -0.6, 0.6) [] {$r$};
			\node at ( 1.05, 2.2) [] {$\ket{O}^{(r)}$};
			\draw [thick] (BL) -- (M1);
			\draw [thick,decorate, decoration=snake] (M1) -- (M2);
			\draw [thick] (M2) -- (T);
			\draw [thick] (BM) -- (M1);
			\draw [thick] (BR) -- (M2);
		\end{tikzpicture} \,
		\quad   \sim \sum_{\ell_{r'}} \int_{\frac{d}{2}}^{\frac{d}{2}+i\nu} \frac{d\Delta_{r'}}{2\pi i} \mu_{r'} W_{1b} (O_1 O_2 O_3 \tilde{O}; O_r O_{r'})           \quad
		\begin{tikzpicture}[anchor=base,baseline,scale=0.8]
			\node [coordinate] (BL) at (-1.5,-1) {};
			\node [coordinate] (BM) at (0, -1) {};
			\node [coordinate] (BR) at ( 1.5,-1) {};
			\node [coordinate] (M1) at ( 0.75, 0) {};
			\node [coordinate] (M2) at ( 0, 1) {};
			\node [coordinate] (T) at ( 0.75, 2) {};
			\node at (-1.5,-1.4) {$1$};
			\node at (0, -1.4) [] {$2$};
			\node at ( 1.5,-1.4) [] {$3$};
			\node at ( 0.6, 0.6) [] {$r'$};
			\node at ( 1.05, 2.2) [] {$\ket{O}^{(r')}$};
			\draw [thick] (BL) -- (T);
			\draw [thick] (BM) -- (M1);
			\draw [thick] (BR) -- (M1);
			\draw [thick, decorate, decoration=snake] (M2) -- (M1);
		\end{tikzpicture} \,
		\end{equation*}
	\caption{A special case of the above: $O_1$, $O_2$, $O_3$, and $O$ are all scalars while the intermediates $O_r$ and $O_{r'}$ have integer spin.}      \label{recoup1a}
	\end{subfigure}
\caption{Transformations mapping different bases for an irrep in the tensor product of three irreps. Scalars are represented with straight lines, spins are depicted with wavy lines.}
\label{fig:Pachner}
\end{figure}

\textbf{Case 1:} Let us begin with the case in fig. \ref{recoup1} wherein we assume that $O_1$ and $O_2$ in the tensor product are scalars while $O_3$ has integer spin. Both the intermediates $O_r$ and $O_{r'}$ (which will be summed over) have integer spin, and the state being extracted corresponding to $O$ is a scalar. We can use \eqref{rCG} followed by \eqref{rCGspin} to obtain,
\beq
\ket{O}^{(r)}  = \frac{1}{C^{(\bullet)}_{3r} C_r} \int dx_1 dx_2 dx_3 \left( \int dx_r \braket{O \tilde{O}_r \tilde{O}_3 }^{(\bullet)} \braket{O_r \tilde{O}_1\tilde{O}_2 } \right) \;\; \ket{O_1 }\otimes \ket{O_2 }\otimes \ket{O_3} ,     \label{recur1}
\eeq 
where we have suppressed the position dependence of the operators in the notation. When we perform the decomposition $\text{Res}^{SO(d)}_{SO(d-1)}\; \rho_{r} = \oplus_{j} \sigma_{r,j} $, one of $\sigma_{r,j}$ is the scalar rep of $SO(d-1)$ and we refer to this with $j=\bullet$. 

Note that $\braket{O_r \tilde{O}_1\tilde{O}_2 }$ is unique and only the $j=\bullet$ component of $\text{Res}^{SO(d)}_{SO(d-1)}\; \rho_{r}$ contributes. This ensures that the only $\braket{O \tilde{O}_r \tilde{O}_3 }^{(j)}$ that is contributing corresponds to $\text{Res}^{SO(d)}_{SO(d-1)}\; \rho_{r} = \bullet$. Therefore we also receive contributions only from $\text{Res}^{SO(d)}_{SO(d-1)}\; \rho_{3} = \bullet$ and we refer to the corresponding 3-pt structure as $\braket{O \tilde{O}_r \tilde{O}_3 }^{(\bullet)}$. This gives us \eqref{recur1}\footnote{Note that $C^{(\bullet)}_{r} = C_{r}$.}.

We could have equally well chosen to take the tensor product in \eqref{recur1} in a different order choosing to expand $\ket{O_2}\otimes\ket{O_3}$ first and gone via a different intermediate state $\ket{O_{r'} }$ (instead of $\ket{O_{r} }$) and we would arrive at,
\beq
\ket{O}^{(r')}  = \frac{1}{C^{(\bullet)}_{3r'} C_{r'}} \int dx_1 dx_2 dx_3 \left( \int dx_{r'} \braket{O \tilde{O}_1 \tilde{O}_{r'}  } \braket{O_{r'} \tilde{O}_2\tilde{O}_3 }^{(\bullet)} \right) \;\; \ket{O_1 }\otimes \ket{O_2 }\otimes \ket{O_3} .     \label{recur2}
\eeq

Let us now define the following ansatz for a change of basis from $\ket{O}^{(r')}$ to the state $\ket{O}^{(r)}$,
\beq
\ket{O}^{(r)} = \sum_{\ell_{r'}} \int_{\frac{d}{2}}^{\frac{d}{2}+i\infty} \frac{d\Delta_{r'}}{2\pi i}\; \mu_{r'} \; W_{1a} (O_1 O_2 O_3 \tilde{O}; O_r O_{r'}) \; \ket{O}^{(r')}   . \label{defW}
\eeq
We can express this diagrammatically as in fig. \ref{fig:Pachner}\footnote{The diagrammatic representation has been adapted from \cite{Gadde:2017sjg}.} with the present case shown in fig. \ref{recoup1}.

To obtain the kernel $W_{1a} (O_1 O_2 O_3 \tilde{O}; O_r O_{r'})$, we can use \eqref{recur1} and \eqref{recur2} in \eqref{defW}, and this gives us the following,
\bea
\frac{1}{C^{(\bullet)}_{3r} C_r}  \int dx_r \braket{O \tilde{O}_r \tilde{O}_3 }^{(\bullet)} \braket{O_r \tilde{O}_1\tilde{O}_2 }  & = & \sum_{\ell_{r'}} \int_{\frac{d}{2}}^{\frac{d}{2}+i\infty} \frac{d\Delta_{r'}}{2\pi i}\; \mu_{r'} \; W_{1a} (O_1 O_2 O_3 \tilde{O}; O_r O_{r'})      \\
			& & \frac{1}{C^{(\bullet)}_{3r'} C_{r'}} \int dx_{r'} \braket{O \tilde{O}_1 \tilde{O}_{r'}  } \braket{O_{r'} \tilde{O}_2\tilde{O}_3 }^{(\bullet)}    .
\eea{appW}
We can now apply the bubble integral \eqref{bubble} once to the above to obtain,
\beq
\frac{1}{C^{(\bullet)}_{3r} C_r}  \int dx_r \braket{O \tilde{O}_r \tilde{O}_3 }^{(\bullet)} \braket{O_r \tilde{O}_1\tilde{O}_2 } \braket{\tilde{O} O_1 O_{r'}  } = \frac{1}{C^{(\bullet)}_{3r'} } W_{1a} (O_1 O_2 O_3 \tilde{O}; O_r O_{r'}) \braket{O_{r'} \tilde{O}_2\tilde{O}_3 }^{(\bullet)}     .    \label{bubbleonce}
\eeq

To get rid of the remaining three-point structure from the r.h.s., we can integrate both sides against $\braket{\tilde{O}_{r'} O_2 O_3 }^{(\bullet)} $ over the remaining position dependence (and quotient out the redundant configurations) to obtain,
\bea
\frac{1}{C^{(\bullet)}_{3r} C_r} \int \frac{dx_r \; dx_{r'} \; dx_1 dx_2 dx_3 \; dx}{\text{Vol } SO(d+1,1)} \braket{O \tilde{O}_r \tilde{O}_3 }^{(\bullet)} \braket{O_r \tilde{O}_1\tilde{O}_2 } \braket{\tilde{O} O_1 O_{r'}  }  \braket{\tilde{O}_{r'} O_2 O_3 }^{(\bullet)}     \\
                         = \frac{1}{C^{(\bullet)}_{3r'} } W_{1a} (O_1 O_2 O_3 \tilde{O}; O_r O_{r'}) \left(\braket{O_{r'} \tilde{O}_2\tilde{O}_3 }^{(\bullet)} , \braket{\tilde{O}_{r'} O_2 O_3 }^{(\bullet)}\right)   .
\eea{W1}

Using \eqref{3pairab} in \eqref{W1}, we obtain the transformation kernel in \eqref{defW} as an integral over a product of three-point functions as follows,
\bea
W_{1a}(O_1 O_2 O_3 \tilde{O}; O_r O_{r'}) = \frac{1}{C^{(\bullet)}_{3r} C_r} \int \frac{dx_r dx_{r'} dx_1 dx_2 dx_3 dx}{\text{Vol } SO(d+1,1)} \; \braket{O \tilde{O}_r \tilde{O}_3 }^{(\bullet)} \braket{O_r \tilde{O}_1\tilde{O}_2 }    \\
			\braket{\tilde{O} O_1 O_{r'}  }  \braket{\tilde{O}_{r'} O_2 O_3 }^{(\bullet)}        . 
\eea{W2}

All the operators in \eqref{W2} are bosonic and hence we can perform a few permutations to put the expression into the following form,
\bea
W_{1a} (O_1 O_2 O_3 \tilde{O}; O_r O_{r'}) = \frac{1}{C^{(\bullet)}_{3r} C_r} \int \frac{dx_r dx_{r'} dx_1 dx_2 dx_3 dx}{\text{Vol } SO(d+1,1)}  \; \braket{\tilde{O}_1\tilde{O}_2 O_r } \braket{\tilde{O}_r \tilde{O}_3 O}^{(\bullet)}   \\
                      \braket{O_3 O_2 \tilde{O}_{r'}}^{(\bullet)}  \braket{O_{r'} O_1 \tilde{O}}         . 
\eea{W2perm}
Comparing \eqref{W2} with \eqref{6jdef}, we see that\footnote{The dashes in superscript in the 6j symbol refer to the fact that the corresponding 3-pt structure is unique, refer to \eqref{6jdef}.},
\beq
W_{1a} (O_1 O_2 O_3 \tilde{O}; O_r O_{r'}) = \frac{1}{C^{(\bullet)}_{3r} C_r} \; \left\{ \begin{array}{ccc} 1 & 2 & \tilde{r}' \\  3 & \tilde{O} & \tilde{r} \end{array} \right\}^{(-\bullet\bullet -)}  . \label{Was6j}
\eeq
which means we can rewrite \eqref{defW} as,
\beq
\ket{O}^{(r)} = \sum_{\ell_{r'}} \int_{\frac{d}{2}}^{\frac{d}{2}+i\infty} \frac{d\Delta_{r'}}{2\pi i}\; \mu_{r'} \; \frac{1}{C^{(\bullet)}_{3r} C_r} \; \left\{ \begin{array}{ccc} 1 & 2 & \tilde{r}' \\  3 & \tilde{O} & \tilde{r} \end{array} \right\}^{(-\bullet\bullet -)} \; \ket{O}^{(r')}   . \label{pen0}
\eeq

Let us also consider a special case now, refer to fig. \ref{recoup1a}, where we set $O_3$ to be a scalar. The transformation kernel is now given by\footnote{We have dropped the tensor structure labels in the 6j symbol in \eqref{Was6j1a} since all the structures involved are unique.},
\beq
W_{1b} (O_1 O_2 O_3 \tilde{O}; O_r O_{r'}) = \frac{1}{ C^2_r} \; \left\{ \begin{array}{ccc} 1 & 2 & \tilde{r}' \\  3 & \tilde{O} & \tilde{r} \end{array} \right\}  . \label{Was6j1a}
\eeq

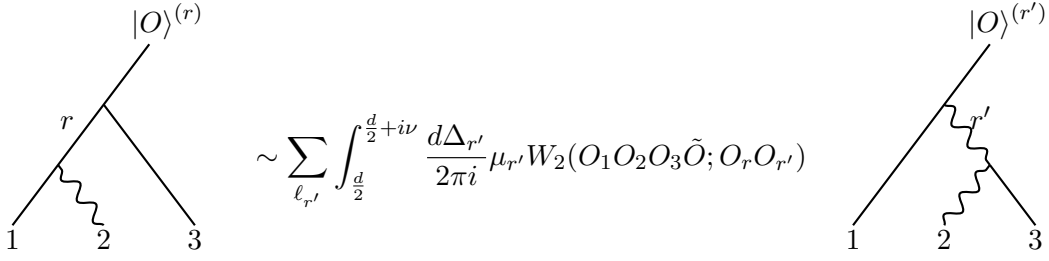
\begin{figure}[h]
		\begin{equation*}
		\begin{tikzpicture}[anchor=base,baseline,scale=0.8]
			\node [coordinate] (BL) at (-1.5,-1) {};
			\node [coordinate] (BM) at (0, -1) {};
			\node [coordinate] (BR) at ( 1.5,-1) {};
			\node [coordinate] (M1) at ( -0.75, 0) {};
			\node [coordinate] (M2) at ( 0, 1) {};
			\node [coordinate] (T) at ( 0.75, 2) {};
			\node at (-1.5,-1.4) {$1$};
			\node at (0, -1.4) [] {$2$};
			\node at ( 1.5,-1.4) [] {$3$};
			\node at ( -0.6, 0.6) [] {$r$};
			\node at ( 1.05, 2.2) [] {$\ket{O}^{(r)}$};
			\draw [thick] (BL) -- (T);
			\draw [thick, decorate, decoration=snake] (BM) -- (M1);
			\draw [thick] (BR) -- (M2);
		\end{tikzpicture} \,
		\quad   \sim \sum_{\ell_{r'}} \int_{\frac{d}{2}}^{\frac{d}{2}+i\nu} \frac{d\Delta_{r'}}{2\pi i} \mu_{r'} W_2 (O_1 O_2 O_3 \tilde{O}; O_r O_{r'})      \quad
		\begin{tikzpicture}[anchor=base,baseline,scale=0.8]
			\node [coordinate] (BL) at (-1.5,-1) {};
			\node [coordinate] (BM) at (0, -1) {};
			\node [coordinate] (BR) at ( 1.5,-1) {};
			\node [coordinate] (M1) at ( 0.75, 0) {};
			\node [coordinate] (M2) at ( 0, 1) {};
			\node [coordinate] (T) at ( 0.75, 2) {};
			\node at (-1.5,-1.4) {$1$};
			\node at (0, -1.4) [] {$2$};
			\node at ( 1.5,-1.4) [] {$3$};
			\node at ( 0.6, 0.6) [] {$r'$};
			\node at ( 1.05, 2.2) [] {$\ket{O}^{(r')}$};
			\draw [thick] (BL) -- (T);
			\draw [thick,decorate,decoration=snake] (BM) -- (M1);
			\draw [thick] (BR) -- (M1);
			\draw [thick, decorate, decoration=snake] (M2) -- (M1);
		\end{tikzpicture} \,
		\end{equation*}
	\caption{$O_2$ and $O_{r'}$ (which is being summed over) have integer spin.  }      \label{recoup2}
\end{figure}
\textbf{Case 2:} Let us now consider $O_2$ and $O_{r'}$ (which is being summed over) to have integer spin and all others including $O_r$ to be scalars. We can write down the analogues of \eqref{recur1} and \eqref{recur2} for this case using \eqref{CGspin} and \eqref{rCGspin}. As shown in fig. \ref{recoup2}, let the kernel for the map from $\ket{O}^{(r')}$ to $\ket{O}^{(r)}$ be $W_2 (O_1 O_2 O_3 \tilde{O}; O_r O_{r'})$ in this case.

The first step is to write the analogue of \eqref{recur1} for this case, and this is essentially identical as the three-point structures involved have at most one integer spin operator and are therefore unique. The factors of $\frac{1}{C_{ijk}}$ are different however and for the spins assumed, we get $\frac{1}{C C_2}$. To get the analogue of \eqref{recur2} note that $\braket{O \tilde{O}_1 \tilde{O}_{r'}  }$ is still unique and only the $j=\bullet$ component of $\text{Res}^{SO(d)}_{SO(d-1)}\; \rho_{r'}$ contributes. This ensures that the only $\braket{O_{r'} \tilde{O}_2\tilde{O}_3 }^{(j)}$ that is relevant is the one that corresponds to $\text{Res}^{SO(d)}_{SO(d-1)}\; \rho_{r'} = \bullet$ and $\text{Res}^{SO(d)}_{SO(d-1)}\; \rho_{2} = \bullet$ i.e. $\braket{O_{r'} \tilde{O}_2\tilde{O}_3 }^{(\bullet)}$. This gives us,
\beq
\ket{O}^{(r')}  = \frac{1}{C_{r'} C^{(\bullet)}_{2r'}} \int dx_1 dx_2 dx_3 \left( \int dx_{r'} \braket{O \tilde{O}_1 \tilde{O}_{r'}  } \braket{O_{r'} \tilde{O}_2\tilde{O}_3 }^{(\bullet)} \right) \;\; \ket{O_1 }\otimes \ket{O_2 }\otimes \ket{O_3} .     \label{recur2case2}
\eeq

To obtain the transformation kernel $W_2$, we can follow the same steps as in \eqref{appW} through \eqref{W1}, and this gives us,
\bea
W_2(O_1 O_2 O_3 \tilde{O}; O_r O_{r'}) = \frac{1}{C C_2} \int \frac{dx_r dx_{r'} dx_1 dx_2 dx_3 dx}{\text{Vol } SO(d+1,1)} \; \braket{O \tilde{O}_r \tilde{O}_3 } \braket{O_r \tilde{O}_1\tilde{O}_2 }  \\ 
		\braket{\tilde{O} O_1 O_{r'}} \braket{\tilde{O}_{r'} O_2 O_3 }^{(\bullet)}         .  
\eea{W2case2}

After performing the necessary permutations to put \eqref{W2case2} into the form \eqref{6jdef}, we obtain,
\beq
W_2(O_1 O_2 O_3 \tilde{O}; O_r O_{r'}) = \frac{1}{C C_2} \; \left\{ \begin{array}{ccc} 1 & 2 & \tilde{r}' \\  3 & \tilde{O} & \tilde{r} \end{array} \right\}^{(--\bullet -)}  . \label{Was6j2}
\eeq

\begin{figure}[h]
		\begin{equation*}
		\begin{tikzpicture}[anchor=base,baseline,scale=0.8]
			\node [coordinate] (BL) at (-1.5,-1) {};
			\node [coordinate] (BM) at (0, -1) {};
			\node [coordinate] (BR) at ( 1.5,-1) {};
			\node [coordinate] (M1) at ( -0.75, 0) {};
			\node [coordinate] (M2) at ( 0, 1) {};
			\node [coordinate] (T) at ( 0.75, 2) {};
			\node at (-1.5,-1.4) {$1$};
			\node at (0, -1.4) [] {$2$};
			\node at ( 1.5,-1.4) [] {$3$};
			\node at ( -0.6, 0.6) [] {$r$};
			\node at ( 1.05, 2.2) [] {$\ket{O}^{(r)}$};
			\draw [thick, decorate, decoration=snake] (BL) -- (M1);
			\draw [thick] (M1) -- (T);
			\draw [thick] (BM) -- (M1);
			\draw [thick] (BR) -- (M2);
		\end{tikzpicture} \,
		\quad   \sim \sum_{\ell_{r'}} \int_{\frac{d}{2}}^{\frac{d}{2}+i\nu} \frac{d\Delta_{r'}}{2\pi i} \mu_{r'} W_3 (O_1 O_2 O_3 \tilde{O}; O_r O_{r'})   \quad
		\begin{tikzpicture}[anchor=base,baseline,scale=0.8]
			\node [coordinate] (BL) at (-1.5,-1) {};
			\node [coordinate] (BM) at (0, -1) {};
			\node [coordinate] (BR) at ( 1.5,-1) {};
			\node [coordinate] (M1) at ( 0.75, 0) {};
			\node [coordinate] (M2) at ( 0, 1) {};
			\node [coordinate] (T) at ( 0.75, 2) {};
			\node at (-1.5,-1.4) {$1$};
			\node at (0, -1.4) [] {$2$};
			\node at ( 1.5,-1.4) [] {$3$};
			\node at ( 0.6, 0.6) [] {$r'$};
			\node at ( 1.05, 2.2) [] {$\ket{O}^{(r')}$};
			\draw [thick] (BM) -- (M1);
			\draw [thick] (BR) -- (M1);
			\draw [thick, decorate, decoration=snake] (BL) -- (M2);
			\draw [thick, decorate, decoration=snake] (M1) -- (M2);
			\draw [thick] (M2) -- (T);
		\end{tikzpicture} \,
		\end{equation*}
	\caption{An exchange of roles between $O_2$ and $O_1$ from the case in fig. \ref{recoup2}.}      \label{recoup3}
\end{figure}
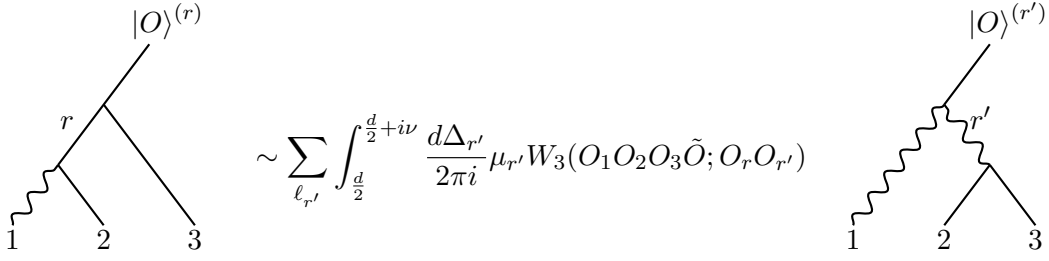

\textbf{Case 3:} Let us now exchange the roles of $O_1$ and $O_2$ from case 2, so that we have $O_1$ and $O_{r'}$ with (integer) spin and the other operators involved as scalars, refer to fig. \ref{recoup3}. This case can be worked out in the same way as the previous one and we obtain:
\beq
W_3(O_1 O_2 O_3 \tilde{O}; O_r O_{r'}) = \frac{1}{C C_1} \; \left\{ \begin{array}{ccc} 1 & 2 & \tilde{r}' \\  3 & \tilde{O} & \tilde{r} \end{array} \right\}^{(---\bullet)}   \label{Was6j3} .
\eeq

\begin{figure}[h]
		\begin{equation*}
		\begin{tikzpicture}[anchor=base,baseline,scale=0.8]
			\node [coordinate] (BL) at (-1.5,-1) {};
			\node [coordinate] (BM) at (0, -1) {};
			\node [coordinate] (BR) at ( 1.5,-1) {};
			\node [coordinate] (M1) at ( -0.75, 0) {};
			\node [coordinate] (M2) at ( 0, 1) {};
			\node [coordinate] (T) at ( 0.75, 2) {};
			\node at (-1.5,-1.4) {$1$};
			\node at (0, -1.4) [] {$2$};
			\node at ( 1.5,-1.4) [] {$3$};
			\node at ( -0.6, 0.6) [] {$r$};
			\node at ( 1.05, 2.2) [] {$\ket{O}^{(r)}$};
			\draw [thick] (BL) -- (M1);
			\draw [thick, decorate, decoration=snake] (M1) -- (M2);
			\draw [thick, decorate, decoration=snake] (M2) -- (T);
			\draw [thick] (BM) -- (M1);
			\draw [thick] (BR) -- (M2);
		\end{tikzpicture} \,
		\quad   \sim \sum_{\ell_{r'}} \int_{\frac{d}{2}}^{\frac{d}{2}+i\nu} \frac{d\Delta_{r'}}{2\pi i} \mu_{r'} W_4 (O_1 O_2 O_3 \tilde{O}; O_r O_{r'})   \quad
		\begin{tikzpicture}[anchor=base,baseline,scale=0.8]
			\node [coordinate] (BL) at (-1.5,-1) {};
			\node [coordinate] (BM) at (0, -1) {};
			\node [coordinate] (BR) at ( 1.5,-1) {};
			\node [coordinate] (M1) at ( 0.75, 0) {};
			\node [coordinate] (M2) at ( 0, 1) {};
			\node [coordinate] (T) at ( 0.75, 2) {};
			\node at (-1.5,-1.4) {$1$};
			\node at (0, -1.4) [] {$2$};
			\node at ( 1.5,-1.4) [] {$3$};
			\node at ( 0.6, 0.6) [] {$r'$};
			\node at ( 1.05, 2.2) [] {$\ket{O}^{(r')}$};
			\draw [thick] (BL) -- (M2);
			\draw [thick] (BM) -- (M1);
			\draw [thick] (BR) -- (M1);
			\draw [thick, decorate, decoration=snake] (M1) -- (M2);
			\draw [thick, decorate, decoration=snake] (M2) -- (T);
		\end{tikzpicture} \,
		\end{equation*}
	\caption{A generalisation of the fig. \ref{recoup1a} to allow the final state $\ket{O}$ have integer spin.}      \label{recoup4}
\end{figure}
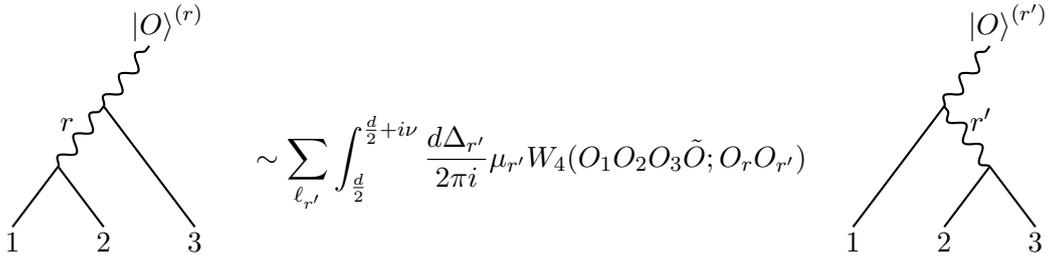

\textbf{Case 4:} This is a generalisation of fig. \ref{recoup1a} to incorporate a final state $\ket{O}$ with integer spin, refer to fig. \ref{recoup4}. The analogue of \eqref{recur1} is as follows,
\beq
\ket{O}^{(r)}_{\bullet}  = \frac{1}{C^{(\bullet)}_{rO}C_r} \int dx_1 dx_2 dx_3 \left( \int dx_r \braket{O \tilde{O}_r \tilde{O}_3 }^{(\bullet)} \braket{O_r \tilde{O}_1\tilde{O}_2 } \right) \;\; \ket{O_1 }\otimes \ket{O_2 }\otimes \ket{O_3} ,     \label{recur1case4}
\eeq
while the analogue of \eqref{recur2} is,
\beq
\ket{O}^{(r')}_{\bullet}  = \frac{1}{C^{(\bullet)}_{r'O}C_{r'}} \int dx_1 dx_2 dx_3 \left( \int dx_{r'} \braket{O \tilde{O}_1 \tilde{O}_{r'}  }^{(\bullet)} \braket{O_{r'} \tilde{O}_2\tilde{O}_3 } \right) \;\; \ket{O_1 }\otimes \ket{O_2 }\otimes \ket{O_3} .     \label{recur2case4}
\eeq

In the same way as before, we can obtain the transformation kernel $W_4$ using the bubble integral and three-point pairings, and then perform some permutations to obtain,
\beq
W_4(O_1 O_2 O_3 \tilde{O}; O_r O_{r'}) = \frac{1}{C^{(\bullet)}_{rO}C_r} \; \left\{ \begin{array}{ccc} 1 & 2 & \tilde{r}' \\  3 & \tilde{O} & \tilde{r} \end{array} \right\}^{(-\bullet -\bullet)}  . \label{Was6j4}
\eeq
As a consistency check, if we set the spin $\ell_O$ to be zero in \eqref{Was6j4}, we get back $W_{1b}$ from \eqref{Was6j1a} as expected. 

Now that we have established that 6j symbols form the kernel of the transforms mapping the different bases in two different schemes of taking tensor product of irreps as in \eqref{recur1} and \eqref{recur2} and the generalisations thereof, the way to reach a solution to the crossing equation in \eqref{posind} is through the pentagon identity for 6j symbols (or Racah coefficients \cite{PhysRev.62.438, PhysRev.63.367}), also known as the Biedenharn-Elliott identity \cite{BiedenharnElliot1, BiedenharnElliot2, BiedenharnElliot3}. We consider this in the next subsection where we present a quick derivation of the identity\footnote{We refer the reader to \cite{Gadde:2017sjg} for a proposal to relate the crossing equation to the pentagon identity and the corresponding analysis for $SU(2)$, from which we draw inspiration for the current project.} and show how it gives us an exact solution to the cosmological bootstrap.

%%%%%%%%%%%%%%%%%%%%%%%%%%%%%%%%%%%%%%%%%%%%%%%%%%%%%%%%%%%%%%%%%%%%%%%%%%%%%%%%%%%%%%%%%%
\subsection{The pentagon identity}
\label{sec:pentagon}
%%%%%%%%%%%%%%%%%%%%%%%%%%%%%%%%%%%%%%%%%%%%%%%%%%%%%%%%%%%%%%%%%%%%%%%%%%%%%%%%%%%%%%%%%%

\begin{figure}[h]
	\begin{subfigure}[b]{0.5\linewidth}
		\begin{tikzpicture}[anchor=base,baseline,scale=0.8]
			\node [coordinate] (v1t1) at (0,0) {};
			\node [coordinate] (v2t1) at (-1,-4) {};
			\node [coordinate] (v3t1) at ( -4,-6) {};
			\node [coordinate] (v4t1) at ( -6,-4) {};
			\draw [thick] (v1t1) -- (v3t1);
			\draw [thick] (v2t1) -- (v3t1);
			\draw [thick] (v4t1) -- (v3t1);
			\draw [thick] (v2t1) -- (v1t1);
			\draw [thick] (v1t1) -- (v4t1);
			\draw [thick, dashed] (v2t1) -- (v4t1);
			\node at (-2.3,-2.2) [rotate=56,anchor=north] {$\mathcolor{red}{1}\equiv 1$};
			\node at (-0.6,-2.2) [rotate=77,anchor=north] {$\mathcolor{red}{4}\equiv 4$};
			\node at (-2.5,-4.85) [rotate=35,anchor=north] {$\mathcolor{red}{6}\equiv r'$};
			\node at (-4,-3.35) [rotate=0,anchor=north] {$\mathcolor{red}{3}\equiv 3$};
			\node at (-4.1,-2) [rotate=35,anchor=north] {$\mathcolor{red}{5}\equiv r$};
			\node at (-5,-5) [rotate=-45,anchor=north] {$\mathcolor{red}{2}\equiv 2$};
			\node [coordinate] (v1t2) at (1,-11) {};
			\node [coordinate] (v2t2) at (-1,-8) {};
			\node [coordinate] (v3t2) at ( -4,-10) {};
			\node [coordinate] (v4t2) at ( -6,-8) {};
			\draw [thick] (v1t2) -- (v3t2);
			\draw [thick] (v2t2) -- (v3t2);
			\draw [thick] (v4t2) -- (v3t2);
			\draw [thick] (v2t2) -- (v1t2);
			\draw [thick] (v2t2) -- (v4t2);
			\draw [thick, dashed] (v1t2) -- (v4t2);
			\node at (-1,-9.45) [rotate=-23,anchor=north] {$\mathcolor{red}{6}\equiv v_1$};
			\node at (-1.7,-10.4) [rotate=-11,anchor=north] {$\mathcolor{red}{4}\equiv v_2$};
			\node at (-3,-8.45) [rotate=35,anchor=north] {$\mathcolor{red}{5}\equiv r'$};
			\node at (-3.9,-7.4) [rotate=0,anchor=north] {$\mathcolor{red}{2}\equiv 3$};
			\node at (0.6,-9.25) [rotate=-55,anchor=north] {$\mathcolor{red}{3}\equiv v_0$};
			\node at (-5,-9) [rotate=-45,anchor=north] {$\mathcolor{red}{1}\equiv 2$};
			\node [coordinate] (v1t3) at (2.5,-0.75) {};
			\node [coordinate] (v2t3) at (3.5,-7.75) {};
			\node [coordinate] (v3t3) at ( -1.5,-6.75) {};
			\node [coordinate] (v4t3) at ( 1.5,-4.75) {};
			\draw [thick] (v1t3) -- (v3t3);
			\draw [thick] (v2t3) -- (v3t3);
			\draw [thick] (v1t3) -- (v2t3);
			\draw [thick, dashed] (v1t3) -- (v4t3);
			\draw [thick, dashed] (v2t3) -- (v4t3);
			\draw [thick, dashed] (v3t3) -- (v4t3);
			\node at (0.1,-3.1) [rotate=56,anchor=north] {$\mathcolor{red}{1}\equiv 1$};
			\node at (3.7,-4.25) [rotate=-80,anchor=north] {$\mathcolor{red}{4}\equiv v_3$};
			\node at (0.7,-7.1) [rotate=-11,anchor=north] {$\mathcolor{red}{6}\equiv v_2$};
			\node at (2.6,-6.25) [rotate=-55,anchor=north] {$\mathcolor{red}{3}\equiv v_0$};
			\node at (1.7,-3.75) [rotate=78,anchor=north] {$\mathcolor{red}{5}\equiv 4$};
			\node at (0.05,-5.55) [rotate=35,anchor=north] {$\mathcolor{red}{2}\equiv r'$};
			\fill[red!100,nearly transparent] (v1t1) -- (v3t1) -- (v2t1) -- cycle;
			\fill[green!70,nearly transparent] (v2t1) -- (v3t1) -- (v4t1) -- cycle;
			\fill[blue!50,nearly transparent] (v1t2) -- (v3t2) -- (v2t2) -- cycle;
			\fill[green!70,nearly transparent] (v2t2) -- (v3t2) -- (v4t2) -- cycle;
			\fill[red!100,nearly transparent] (v1t3) -- (v3t3) -- (v4t3) -- cycle;
			\fill[blue!50,nearly transparent] (v2t3) -- (v3t3) -- (v4t3) -- cycle;
		\end{tikzpicture} 
	\caption{Gluing three tetrahedra on faces sharing a common edge to form a hexahedron with the common edge now interior to said hexahedron.}      \label{tetraglue1}
	\end{subfigure}	\hfill
	\begin{subfigure}[b]{0.4\linewidth}
	\begin{tikzpicture}[anchor=base,baseline,scale=0.9]
			\node [coordinate] (v1t1) at (0,2) {};
			\node [coordinate] (v2t1) at (-2.5,-1.5) {};
			\node [coordinate] (v3t1) at ( 0,-3) {};
			\node [coordinate] (v4t1) at ( 2.5,-1.5) {};
			\draw [thick] (v1t1) -- (v3t1);
			\draw [thick] (v2t1) -- (v3t1);
			\draw [thick] (v4t1) -- (v3t1);
			\draw [thick] (v2t1) -- (v1t1);
			\draw [thick] (v1t1) -- (v4t1);
			\draw [thick, dashed] (v2t1) -- (v4t1);
			\node at (-1.7,0.75) [rotate=57,anchor=north] {$\mathcolor{red}{3}\equiv v_0$};
			\node at (1.7,0.75) [rotate=-57,anchor=north] {$\mathcolor{red}{2}\equiv 3$};
			\node at (0,-0.4) [rotate=-90,anchor=north] {$\mathcolor{red}{5}\equiv 4$};
			\node at (1,-0.9) [rotate=0,anchor=north] {$\mathcolor{red}{6}\equiv v_1$};
			\node at (-1.5,-2.1) [rotate=-30,anchor=north] {$\mathcolor{red}{4}\equiv v_3$};
			\node at (1.5,-2.1) [rotate=32,anchor=north] {$\mathcolor{red}{1}\equiv r$};
			\node [coordinate] (v1t2) at (0,-7) {};
			\node [coordinate] (v2t2) at (-2.5,-3.5) {};
			\node [coordinate] (v3t2) at ( 0,-5) {};
			\node [coordinate] (v4t2) at ( 2.5,-3.5) {};
			\draw [thick] (v1t2) -- (v3t2);
			\draw [thick] (v2t2) -- (v3t2);
			\draw [thick] (v4t2) -- (v3t2);
			\draw [thick] (v2t2) -- (v1t2);
			\draw [thick] (v2t2) -- (v4t2);
			\draw [thick] (v1t2) -- (v4t2);
			\node at (-1.4,-5) [rotate=-54,anchor=north] {$\mathcolor{red}{6}\equiv v_2$};
			\node at (1.4,-5) [rotate=54,anchor=north] {$\mathcolor{red}{2}\equiv 2$};
			\node at (0,-5.6) [rotate=-90,anchor=north] {$\mathcolor{red}{1}\equiv 1$};
			\node at (0,-2.9) [rotate=0,anchor=north] {$\mathcolor{red}{3}\equiv v_1$};
			\node at (-1.05,-3.75) [rotate=-28,anchor=north] {$\mathcolor{red}{4}\equiv v_3$};
			\node at (0.8,-3.8) [rotate=32,anchor=north] {$\mathcolor{red}{5}\equiv r$};
			\fill[gray!100,nearly transparent] (v2t1) -- (v3t1) -- (v4t1) -- cycle;
			\fill[gray!100,nearly transparent] (v2t2) -- (v3t2) -- (v4t2) -- cycle;
		\end{tikzpicture} 
	\caption{An alternative tessellation of a hexahedron in terms of tetrahedra: gluing two tetrahedra on a common face.}      \label{tetraglue2}
	\end{subfigure}
\caption{Different tessellations of a hexahedron in terms of tetrahedra. The faces being glued are shaded in the same colour. The labels on the edges coloured in red indicate the corresponding position in the 6j symbol, while the labels in black indicate the operator associated to that position in the 6j symbols in \eqref{nettrans1} and \eqref{nettrans2} resp. }
\label{fig:lat}
\end{figure}
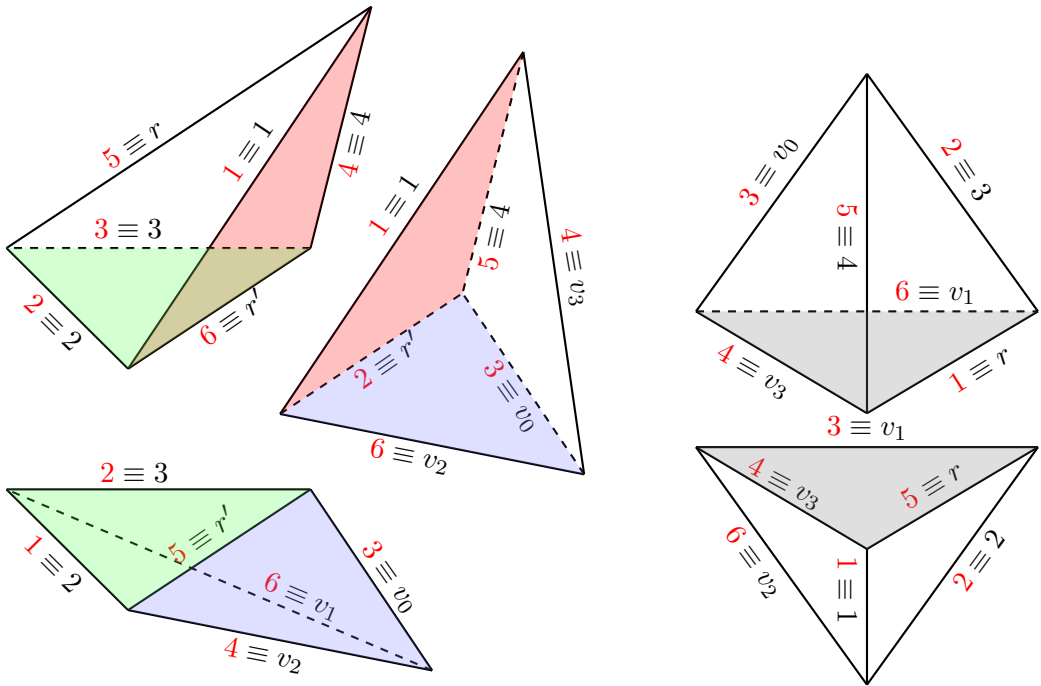

A pentagon identity is a relation that is fundamentally an equivalence of the two different ways of tessellating a hexahedron with tetrahedra. One can either glue three tetrahedra on faces sharing a common edge that is interior to the newly formed hexagon as depicted in fig. \ref{tetraglue1}, or equivalently one can glue two tetrahedra on a common face as shown in fig. \ref{tetraglue2}. 

The pentagon identity for 6j symbols is a consequence of the associativity of the tensor product of irreps of states in a conformal theory and can be shown using \eqref{pen0} (and the analogous equations that we discussed subsequently)  recursively. The idea is to consider a transform between two different bases associated with a different order of tensoring the four irreps, and then to show the equivalence of two different schemes wherein we compose the transformations in \eqref{Was6j}, \eqref{Was6j1a}, \eqref{Was6j2}, \eqref{Was6j3}, \eqref{Was6j4} to produce said transform. 

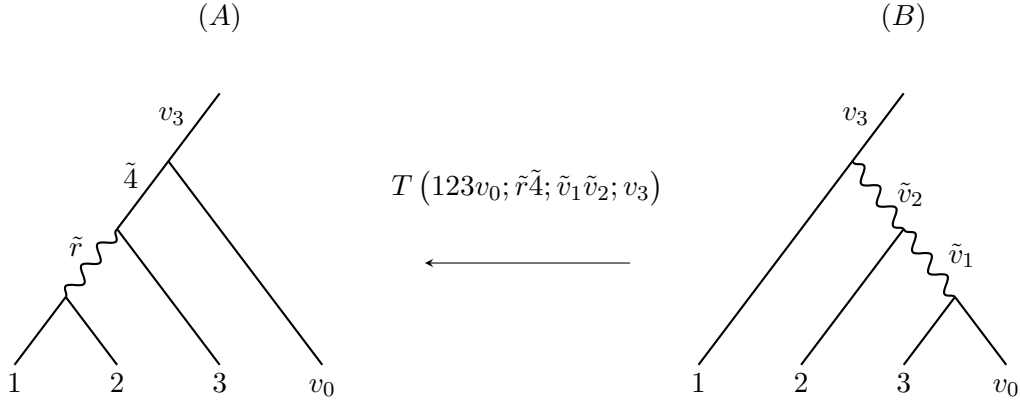
\begin{figure}
\begin{tikzpicture}[anchor=base,baseline, scale=0.9]
	\node [coordinate] (f1_b1) at (-1.5,-1) {};
	\node [coordinate] (f1_b2) at (0, -1) {};
	\node [coordinate] (f1_b3) at ( 1.5,-1) {};
	\node [coordinate] (f1_b4) at ( 3, -1) {};
	\node [coordinate] (f1_f1) at ( -0.75, 0) {};
	\node [coordinate] (f1_f2) at ( 0.75, 0) {};
	\node [coordinate] (f1_f3) at ( 2.25, 0) {};
	\node [coordinate] (f1_s1) at ( 0, 1) {};
	\node [coordinate] (f1_s2) at ( 1.5, 1) {};
	\node [coordinate] (f1_th1) at ( 0.75, 2) {};
	\node [coordinate] (f1_t) at ( 1.5, 3) {};
	\draw [thick] (f1_b1) -- (f1_f1);
	\draw [thick, decorate, decoration=snake] (f1_f1) -- (f1_s1);
	\draw [thick] (f1_s1) -- (f1_t);
	\draw [thick] (f1_b2) -- (f1_f1);
	\draw [thick] (f1_b3) -- (f1_s1);
	\draw [thick] (f1_b4) -- (f1_th1);	
	\node at (-1.5,-1.4) {$1$};
	\node at (0, -1.4) [] {$2$};
	\node at (1.5, -1.4) [] {$3$};
	\node at ( 3,-1.4) [] {$v_0$};
	\node at ( -0.6, 0.6) [] {$\tilde{r}$};
	\node at ( 0.2, 1.6) [] {$\tilde{4}$};
	\node at ( 0.8, 2.6) [] {$v_3$};
	\node at (6, 1.5) [] {$T\left(123v_0; \tilde{r}\tilde{4}; \tilde{v}_1 \tilde{v}_2 ; v_3\right)$};
	%\node at (6, 0.5) [] {$\xrightarrow{\hspace*{3cm}}$};
	\draw[stealth-] (4.5,0.5) -- (7.5,0.5);
	\node [coordinate] (f2_b1) at (8.5,-1) {};
	\node [coordinate] (f2_b2) at (10, -1) {};
	\node [coordinate] (f2_b3) at ( 11.5,-1) {};
	\node [coordinate] (f2_b4) at ( 13, -1) {};
	\node [coordinate] (f2_f1) at ( 9.25, 0) {};
	\node [coordinate] (f2_f2) at ( 10.75, 0) {};
	\node [coordinate] (f2_f3) at ( 12.25, 0) {};
	\node [coordinate] (f2_s1) at ( 10, 1) {};
	\node [coordinate] (f2_s2) at ( 11.5, 1) {};
	\node [coordinate] (f2_th1) at ( 10.75, 2) {};
	\node [coordinate] (f2_t) at ( 11.5, 3) {};
	\draw [thick] (f2_b1) -- (f2_t);
	\draw [thick] (f2_b2) -- (f2_s2);
	\draw [thick] (f2_b3) -- (f2_f3);
	\draw [thick] (f2_b4) -- (f2_f3);
	\draw [thick, decorate, decoration=snake] (f2_f3) -- (f2_s2);
	\draw [thick, decorate, decoration=snake] (f2_s2) -- (f2_th1);	
	\node at (8.5,-1.4) {$1$};
	\node at (10, -1.4) [] {$2$};
	\node at (11.5, -1.4) [] {$3$};
	\node at (13, -1.4) [] {$v_0$};
	\node at ( 12.35, 0.5) [] {$\tilde{v}_1$};
	\node at ( 11.6, 1.4) [] {$\tilde{v}_2$};
	\node at ( 10.8, 2.6) [] {$v_3$};
	\node at ( 1.5, 4) [] {$(A)$};
	\node at ( 11.5, 4) [] {$(B)$};
\end{tikzpicture}
\caption{$T\left(123v_0; \tilde{r}\tilde{4}; \tilde{v}_1 \tilde{v}_2 ; v_3\right)$ transforms $\ket{v_3}$ from the basis $B$ to the basis $A$ associated with the different orders in which we take the tensor product of $\ket{1}$, $\ket{2}$, $\ket{3}$, and $\ket{v_0}$.}
\label{fig:diffbases}	
\end{figure}

Consider a tensor product of the irreps of $\ket{1}$, $\ket{2}$, $\ket{3}$, and $\ket{v_0}$, and $\ket{v_3}$ appearing in the decomposition of this tensor product. We can carry out the tensoring in the order $\left(\left(\ket{1} \otimes \ket{2}\right) \otimes \ket{3}\right) \otimes \ket{v_0}$ (as shown in fig. \ref{fig:diffbases} on the left) and say we extract $\ket{v_3}^{(A)}$ from the tensor product. We can also carry out the tensoring in the order $\ket{1} \otimes \left( \ket{2} \otimes \left( \ket{3} \otimes \ket{v_0} \right)\right)$ (as shown in fig. \ref{fig:diffbases} on the right), and let us denote the extracted irrep in this case as $\ket{v_3}^{(B)}$. Let's say the transform from $\ket{v_3}^{(B)}$ to $\ket{v_3}^{(A)}$ is $T\left(123v_0; \tilde{r}\tilde{4}; \tilde{v}_1 \tilde{v}_2 ; v_3\right)$. We are free to choose $\ket{1}$, $\ket{2}$, $\ket{3}$, $\ket{v_0}$, $\ket{\tilde{r}}$, $\ket{\tilde{4}}$, and $\ket{v_3}$ as we wish and we choose all of these except $\ket{r}$ to be scalars (for simplicity).
\beq
\ket{v_3}^{(A)} = \sum_{\ell_{v_2}, \ell_{v_1}} \int_{\frac{d}{2}}^{\frac{d}{2}+i\infty} \frac{d\tilde{\Delta}_{v_2}}{2\pi i} \mu_{v_2} \int_{\frac{d}{2}}^{\frac{d}{2}+i\infty} \frac{d\tilde{\Delta}_{v_1}}{2\pi i} \mu_{v_1}   \;  T\left(123v_0; \tilde{r}\tilde{4}; \tilde{v}_1 \tilde{v}_2 ; v_3\right) \; \ket{v_3}^{(B)}  \label{nettranspach}.
\eeq

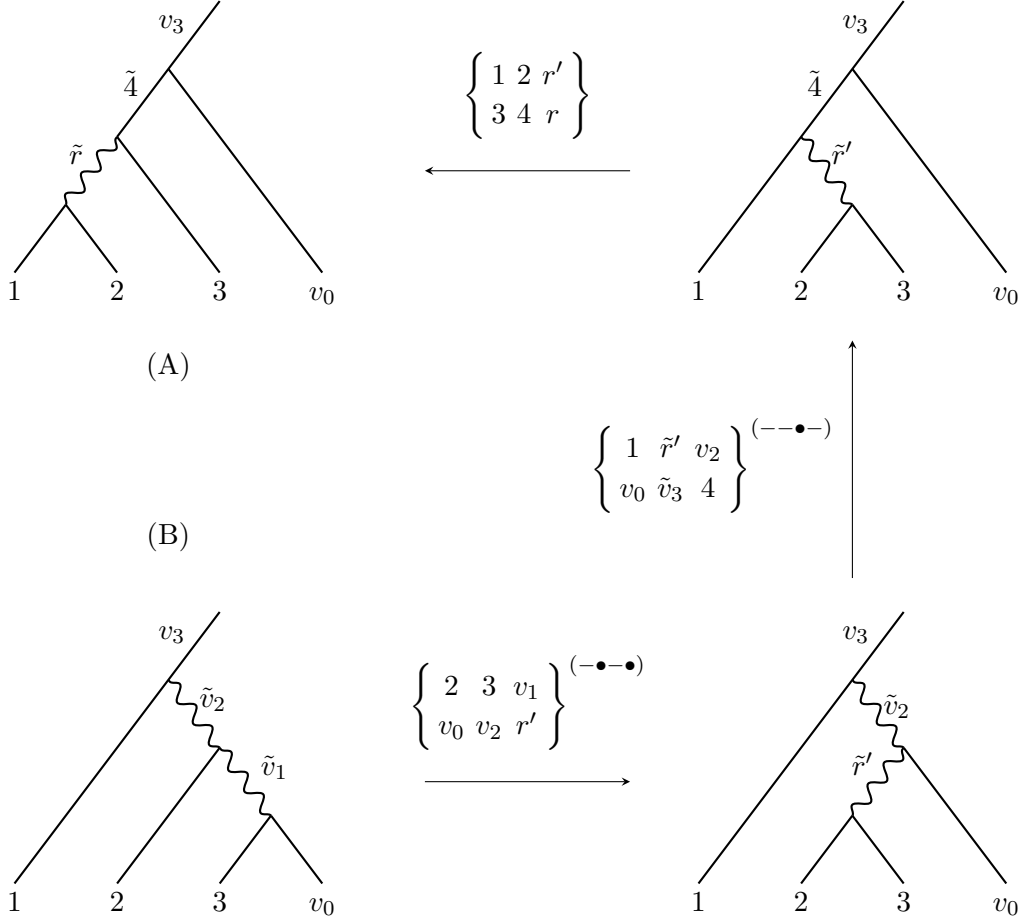
\begin{figure}[h]
\begin{tikzpicture}[anchor=base,baseline, scale=0.9]
	\node [coordinate] (f1_b1) at (-1.5,-1) {};
	\node [coordinate] (f1_b2) at (0, -1) {};
	\node [coordinate] (f1_b3) at ( 1.5,-1) {};
	\node [coordinate] (f1_b4) at ( 3, -1) {};
	\node [coordinate] (f1_f1) at ( -0.75, 0) {};
	\node [coordinate] (f1_f2) at ( 0.75, 0) {};
	\node [coordinate] (f1_f3) at ( 2.25, 0) {};
	\node [coordinate] (f1_s1) at ( 0, 1) {};
	\node [coordinate] (f1_s2) at ( 1.5, 1) {};
	\node [coordinate] (f1_th1) at ( 0.75, 2) {};
	\node [coordinate] (f1_t) at ( 1.5, 3) {};
	\draw [thick] (f1_b1) -- (f1_f1);
	\draw [thick, decorate, decoration=snake] (f1_f1) -- (f1_s1);
	\draw [thick] (f1_s1) -- (f1_t);
	\draw [thick] (f1_b2) -- (f1_f1);
	\draw [thick] (f1_b3) -- (f1_s1);
	\draw [thick] (f1_b4) -- (f1_th1);	
	\node at (-1.5,-1.4) {$1$};
	\node at (0, -1.4) [] {$2$};
	\node at (1.5, -1.4) [] {$3$};
	\node at ( 3,-1.4) [] {$v_0$};
	\node at ( -0.6, 0.6) [] {$\tilde{r}$};
	\node at ( 0.2, 1.6) [] {$\tilde{4}$};
	\node at ( 0.8, 2.6) [] {$v_3$};
	\node at (6, 1.5) [] {$\left\{ \begin{array}{ccc} 1 & 2 & r' \\  3 & 4 & r \end{array} \right\} $};
	%\node at (6, 0.5) [] {$\xrightarrow{\hspace*{3cm}}$};
	\draw[stealth-] (4.5,0.5) -- (7.5,0.5);
	\node [coordinate] (f2_b1) at (8.5,-1) {};
	\node [coordinate] (f2_b2) at (10, -1) {};
	\node [coordinate] (f2_b3) at ( 11.5,-1) {};
	\node [coordinate] (f2_b4) at ( 13, -1) {};
	\node [coordinate] (f2_f1) at ( 9.25, 0) {};
	\node [coordinate] (f2_f2) at ( 10.75, 0) {};
	\node [coordinate] (f2_f3) at ( 12.25, 0) {};
	\node [coordinate] (f2_s1) at ( 10, 1) {};
	\node [coordinate] (f2_s2) at ( 11.5, 1) {};
	\node [coordinate] (f2_th1) at ( 10.75, 2) {};
	\node [coordinate] (f2_t) at ( 11.5, 3) {};
	\draw [thick] (f2_b1) -- (f2_t);
	\draw [thick] (f2_b2) -- (f2_f2);
	\draw [thick] (f2_b3) -- (f2_f2);
	\draw [thick, decorate, decoration=snake] (f2_f2) -- (f2_s1);
	\draw [thick] (f2_b4) -- (f2_th1);	
	\node at (8.5,-1.4) {$1$};
	\node at (10, -1.4) [] {$2$};
	\node at (11.5, -1.4) [] {$3$};
	\node at (13, -1.4) [] {$v_0$};
	\node at ( 10.6, 0.6) [] {$\tilde{r}'$};
	\node at ( 10.2, 1.6) [] {$\tilde{4}$};
	\node at ( 10.8, 2.6) [] {$v_3$};
	\node at (8.7, -4) [] {$\left\{ \begin{array}{ccc} 1 & \tilde{r}' & v_2 \\  v_0 & \tilde{v}_3 & 4 \end{array} \right\}^{(--\bullet -)} $};
	%\node at (10.75, -4) [] {$\xdownarrow{1.8cm}$};
	\draw[stealth-] (10.75,-2) -- (10.75,-5.5);
	\node [coordinate] (f3_b1) at (8.5,-10) {};
	\node [coordinate] (f3_b2) at (10, -10) {};
	\node [coordinate] (f3_b3) at ( 11.5,-10) {};
	\node [coordinate] (f3_b4) at ( 13, -10) {};
	\node [coordinate] (f3_f1) at ( 9.25, -9) {};
	\node [coordinate] (f3_f2) at ( 10.75, -9) {};
	\node [coordinate] (f3_f3) at ( 12.25, -9) {};
	\node [coordinate] (f3_s1) at ( 10, -8) {};
	\node [coordinate] (f3_s2) at ( 11.5, -8) {};
	\node [coordinate] (f3_th1) at ( 10.75, -7) {};
	\node [coordinate] (f3_t) at ( 11.5, -6) {};
	\draw [thick] (f3_b1) -- (f3_t);
	\draw [thick] (f3_b2) -- (f3_f2);
	\draw [thick, decorate, decoration=snake] (f3_f2) -- (f3_s2);
	\draw [thick] (f3_b3) -- (f3_f2);
	\draw [thick] (f3_b4) -- (f3_s2);
	\draw [thick, decorate, decoration=snake] (f3_s2) -- (f3_th1);	
	\node at (8.5,-10.4) {$1$};
	\node at (10, -10.4) [] {$2$};
	\node at (11.5, -10.4) [] {$3$};
	\node at (13, -10.4) [] {$v_0$};
	\node at ( 10.9, -8.4) [] {$\tilde{r}'$};
	\node at ( 11.4, -7.5) [] {$\tilde{v}_2$};
	\node at ( 10.8, -6.4) [] {$v_3$};
	\node at (6, -7.5) [] {$\left\{ \begin{array}{ccc} 2 & 3 & v_1 \\  v_0 & v_2 & r' \end{array} \right\}^{(-\bullet -\bullet)}  $};
	\draw[stealth-] (7.5,-8.5) -- (4.5,-8.5) ;
	\node [coordinate] (f4_b1) at (-1.5,-10) {};
	\node [coordinate] (f4_b2) at (0, -10) {};
	\node [coordinate] (f4_b3) at ( 1.5,-10) {};
	\node [coordinate] (f4_b4) at ( 3, -10) {};
	\node [coordinate] (f4_f1) at ( -0.75, -9) {};
	\node [coordinate] (f4_f2) at ( 0.75, -9) {};
	\node [coordinate] (f4_f3) at ( 2.25, -9) {};
	\node [coordinate] (f4_s1) at ( 0, -8) {};
	\node [coordinate] (f4_s2) at ( 1.5, -8) {};
	\node [coordinate] (f4_th1) at ( 0.75, -7) {};
	\node [coordinate] (f4_t) at ( 1.5, -6) {};
	\draw [thick] (f4_b1) -- (f4_t);
	\draw [thick] (f4_b2) -- (f4_s2);
	\draw [thick] (f4_b3) -- (f4_f3);
	\draw [thick] (f4_b4) -- (f4_f3);
	\draw [thick, decorate, decoration=snake] (f4_f3) -- (f4_s2);
	\draw [thick, decorate, decoration=snake] (f4_s2) -- (f4_th1);	
	\node at (-1.5,-10.4) {$1$};
	\node at (0, -10.4) [] {$2$};
	\node at (1.5, -10.4) [] {$3$};
	\node at ( 3,-10.4) [] {$v_0$};
	\node at ( 2.3, -8.4) [] {$\tilde{v}_1$};
	\node at ( 1.4, -7.4) [] {$\tilde{v}_2$};
	\node at ( 0.8, -6.4) [] {$v_3$};
	\node at ( 0.75, -2.5) [] {(A)};
	\node at ( 0.75, -5) [] {(B)};
\end{tikzpicture}
\caption{A composition of transformations to map $\ket{v_3}^{(B)}$ to $\ket{v_3}^{(A)}$ going through two intermediate bases for $v_3$. At each step, the transformation involves an integral over the principal series with the kernel being proportional to a 6j symbol. We have shown the appropriate 6j symbol in the figure while suppressing the other details for brevity.}
\label{fig:moves1}	
\end{figure}

$T\left(123v_0; \tilde{r}\tilde{4}; \tilde{v}_1 \tilde{v}_2 ; v_3\right)$ can be composed from the transforms discussed in the previous section in two different ways. Consider for example the steps depicted in fig. \eqref{fig:moves1}. In this case we use the transforms $W_4$ from \eqref{Was6j4}, $W_2$ from \eqref{Was6j2}, and $W_{1b}$ from \eqref{Was6j1a}, and the net transform is given by\footnote{Note that $C_O = C_{\tilde{O}}$.},
\beq
T = \sum_{\ell_{r'}} \int_{\frac{d}{2}}^{\frac{d}{2}+i\infty} \frac{d\tilde{\Delta}_{r'}}{2\pi i}\; \mu_{r'} \; \frac{1}{C C^2_r C^2_{r'} C^{(\bullet)}_{v_2 r'} }    \left\{ \begin{array}{ccc} 1 & 2 & r' \\  3 & 4 & r \end{array} \right\}  \; \left\{ \begin{array}{ccc} 1 & \tilde{r}' & v_2 \\  v_0 & \tilde{v}_3 & 4 \end{array} \right\}^{(--\bullet -)} \; \left\{ \begin{array}{ccc} 2 & 3 & v_1 \\  v_0 & v_2 & r' \end{array} \right\}^{(-\bullet -\bullet)}       .          \label{nettrans1}
\eeq

\begin{figure}
\begin{tikzpicture}[anchor=base,baseline, scale=0.9]
	\node [coordinate] (f1_b1) at (-1.5,-1) {};
	\node [coordinate] (f1_b2) at (0, -1) {};
	\node [coordinate] (f1_b3) at ( 1.5,-1) {};
	\node [coordinate] (f1_b4) at ( 3, -1) {};
	\node [coordinate] (f1_f1) at ( -0.75, 0) {};
	\node [coordinate] (f1_f2) at ( 0.75, 0) {};
	\node [coordinate] (f1_f3) at ( 2.25, 0) {};
	\node [coordinate] (f1_s1) at ( 0, 1) {};
	\node [coordinate] (f1_s2) at ( 1.5, 1) {};
	\node [coordinate] (f1_th1) at ( 0.75, 2) {};
	\node [coordinate] (f1_t) at ( 1.5, 3) {};
	\draw [thick] (f1_b1) -- (f1_f1);
	\draw [thick, decorate, decoration=snake] (f1_f1) -- (f1_s1);
	\draw [thick] (f1_s1) -- (f1_t);
	\draw [thick] (f1_b2) -- (f1_f1);
	\draw [thick] (f1_b3) -- (f1_s1);
	\draw [thick] (f1_b4) -- (f1_th1);	
	\node at (-1.5,-1.4) {$1$};
	\node at (0, -1.4) [] {$2$};
	\node at (1.5, -1.4) [] {$3$};
	\node at ( 3,-1.4) [] {$v_0$};
	\node at ( -0.6, 0.6) [] {$\tilde{r}$};
	\node at ( 0.2, 1.6) [] {$\tilde{4}$};
	\node at ( 0.8, 2.6) [] {$v_3$};
	\node at (6, 1.5) [] {$\left\{ \begin{array}{ccc} \tilde{r} & 3 & v_1 \\  v_0 & \tilde{v}_3 & 4 \end{array} \right\}^{(---\bullet)} $};
	%\node at (6, 0.5) [] {$\xrightarrow{\hspace*{3cm}}$};
	\draw[stealth-] (4.5,0.5) -- (7.5,0.5);
	\node [coordinate] (f2_b1) at (8.5,-1) {};
	\node [coordinate] (f2_b2) at (10, -1) {};
	\node [coordinate] (f2_b3) at ( 11.5,-1) {};
	\node [coordinate] (f2_b4) at ( 13, -1) {};
	\node [coordinate] (f2_f1) at ( 9.25, 0) {};
	\node [coordinate] (f2_f2) at ( 10.75, 0) {};
	\node [coordinate] (f2_f3) at ( 12.25, 0) {};
	\node [coordinate] (f2_s1) at ( 10, 1) {};
	\node [coordinate] (f2_s2) at ( 11.5, 1) {};
	\node [coordinate] (f2_th1) at ( 10.75, 2) {};
	\node [coordinate] (f2_t) at ( 11.5, 3) {};
	\draw [thick] (f2_b1) -- (f2_f1);
	\draw [thick, decorate, decoration=snake] (f2_f1) -- (f2_th1);
	\draw [thick] (f2_th1) -- (f2_t);
	\draw [thick] (f2_b2) -- (f2_f1);
	\draw [thick] (f2_b3) -- (f2_f3);
	\draw [thick] (f2_b4) -- (f2_f3);
	\draw [thick, decorate, decoration=snake] (f2_f3) -- (f2_th1);	
	\node at (8.5,-1.4) {$1$};
	\node at (10, -1.4) [] {$2$};
	\node at (11.5, -1.4) [] {$3$};
	\node at (13, -1.4) [] {$v_0$};
	\node at ( 12.2, 0.6) [] {$\tilde{v}_1$};
	\node at ( 9.3, 0.6) [] {$\tilde{r}$};
	\node at ( 10.8, 2.6) [] {$v_3$};
	\node at (10.8, -5.25) [] {$\left\{ \begin{array}{ccc} 1 & 2 & v_2 \\  \tilde{v}_1 & \tilde{v}_3 & r \end{array} \right\}^{(-\bullet\bullet -)} $};
	%\node at (6, 0.5) [] {$\xrightarrow{\hspace*{3cm}}$};
	\draw[stealth-] (10.75,-2.5) -- (7.5, -6.75);
	\node [coordinate] (f4_b1) at (3,-10) {};
	\node [coordinate] (f4_b2) at (4.5, -10) {};
	\node [coordinate] (f4_b3) at ( 6,-10) {};
	\node [coordinate] (f4_b4) at ( 7.5, -10) {};
	\node [coordinate] (f4_f1) at ( 3.75, -9) {};
	\node [coordinate] (f4_f2) at ( 5.25, -9) {};
	\node [coordinate] (f4_f3) at ( 6.75, -9) {};
	\node [coordinate] (f4_s1) at ( 4.5, -8) {};
	\node [coordinate] (f4_s2) at ( 6, -8) {};
	\node [coordinate] (f4_th1) at ( 5.25, -7) {};
	\node [coordinate] (f4_t) at ( 6, -6) {};
	\draw [thick] (f4_b1) -- (f4_t);
	\draw [thick] (f4_b2) -- (f4_s2);
	\draw [thick] (f4_b3) -- (f4_f3);
	\draw [thick] (f4_b4) -- (f4_f3);
	\draw [thick, decorate, decoration=snake] (f4_f3) -- (f4_s2);
	\draw [thick, decorate, decoration=snake] (f4_s2) -- (f4_th1);	
	\node at (3,-10.4) {$1$};
	\node at (4.5, -10.4) [] {$2$};
	\node at (6, -10.4) [] {$3$};
	\node at ( 7.5,-10.4) [] {$v_0$};
	\node at ( 6.8, -8.4) [] {$\tilde{v}_1$};
	\node at ( 5.9, -7.4) [] {$\tilde{v}_2$};
	\node at ( 5.3, -6.4) [] {$v_3$};
	\node at ( 0.75, -2.5) [] {(A)};
	\node at ( 5.25, -5) [] {(B)};
\end{tikzpicture}
\caption{A second scheme of transformations to map $\ket{v_3}^{(B)}$ to $\ket{v_3}^{(A)}$.}
\label{fig:moves2}	
\end{figure}
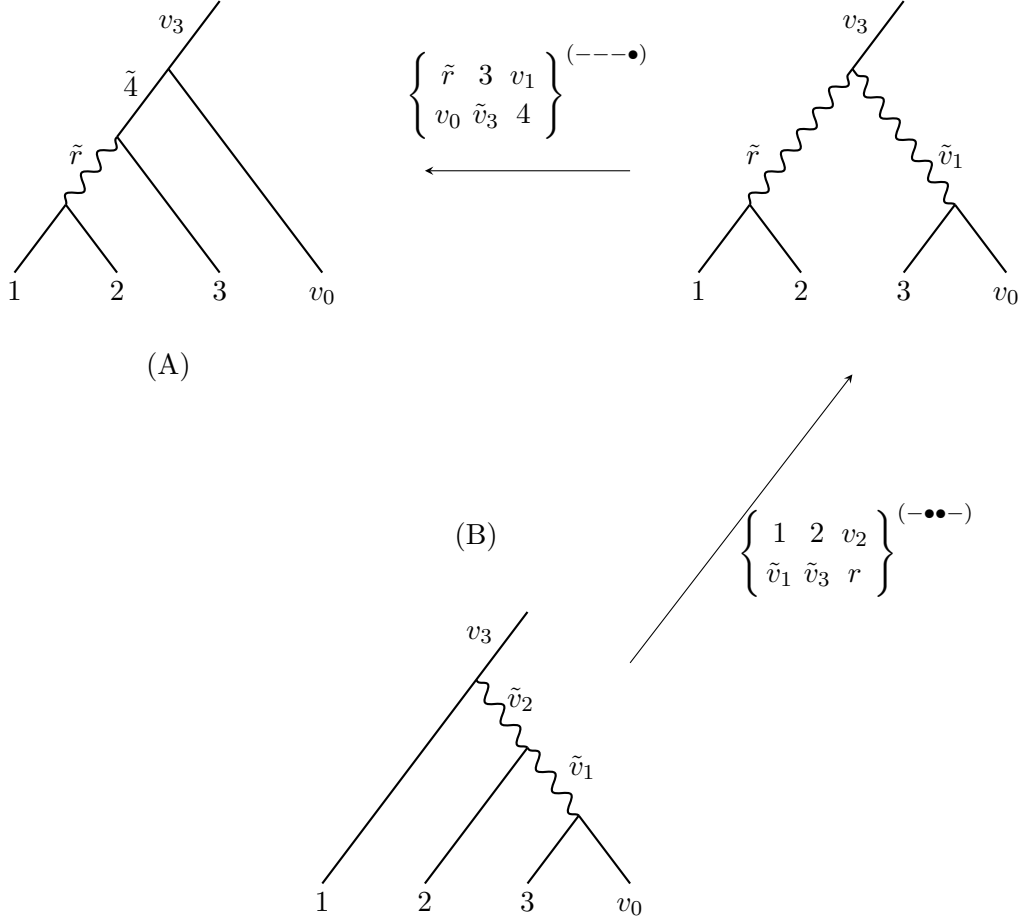

We could, however, also arrive from $\ket{v_3}^{(B)}$ to $\ket{v_3}^{(A)}$ using a different scheme of transformations as depicted in fig. \ref{fig:moves2}. In this case we use $W_{1a}$ from \eqref{Was6j} and $W_3$ from \eqref{Was6j3} to obtain,
\beq
T = \frac{1}{C C^2_r  C_{r v_1}^{(\bullet)}}  \left\{ \begin{array}{ccc} \tilde{r} & 3 & v_1 \\  v_0 & \tilde{v}_3 & 4 \end{array} \right\}^{(---\bullet)} \; \left\{ \begin{array}{ccc} 1 & 2 & v_2 \\  \tilde{v}_1 & \tilde{v}_3 & r \end{array} \right\}^{(-\bullet\bullet -)}                  .   \label{nettrans2}
\eeq

We can now choose to look at the case when $v_1$ and $v_2$ are both scalars in the equality between \eqref{nettrans1} and \eqref{nettrans2}, which gives us,
\bea
\sum_{\ell_{r'}} \int_{\frac{d}{2}}^{\frac{d}{2}+i\infty} \frac{d\tilde{\Delta}_{r'}}{2\pi i}\; \mu_{r'} \; \frac{1}{C^3_{r'}}    \left\{ \begin{array}{ccc} 1 & 2 & r' \\  3 & 4 & r \end{array} \right\}  \; \left\{ \begin{array}{ccc} 1 & \tilde{r}' & v_2 \\  v_0 & \tilde{v}_3 & 4 \end{array} \right\} \; \left\{ \begin{array}{ccc} 2 & 3 & v_1 \\  v_0 & v_2 & r' \end{array} \right\}    \\
     = \frac{1}{C_{r}}  \left\{ \begin{array}{ccc} \tilde{r} & 3 & v_1 \\  v_0 & \tilde{v}_3 & 4 \end{array} \right\} \; \left\{ \begin{array}{ccc} 1 & 2 & v_2 \\  \tilde{v}_1 & \tilde{v}_3 & r \end{array} \right\}          .   
\eea{pentaspecial}
Thus we have derived, in \eqref{pentaspecial}, a pentagon identity for the 6j symbols of an $SO(d+1,1)$ symmetric theory, see fig. \ref{fig:lat}. We shall now show how \eqref{pentaspecial} can be transformed into the crossing equation in \eqref{posind} which gives us a solution for the three-point function coefficients $K_{ijk}$ in terms of 6j symbols.

%%%%%%%%%%%%%%%%%%%%%%%%%%%%%%%%%%%%%%%%%%%%%%%%%%%%%%%%%%%%%%%%%%%%%%%%%%%%%%%%%%%%%%%%%%
\subsection{The crossing equation}
\label{sec:crossoln}
%%%%%%%%%%%%%%%%%%%%%%%%%%%%%%%%%%%%%%%%%%%%%%%%%%%%%%%%%%%%%%%%%%%%%%%%%%%%%%%%%%%%%%%%%%

Using the tetrahedral symmetries of the 6j symbols as shown in \eqref{tetra}, we can put \eqref{pentaspecial} into the following form,
\bea
\sum_{\ell_{r'}} \int_{\frac{d}{2}}^{\frac{d}{2}+i\infty} \frac{d\tilde{\Delta}_{r'}}{2\pi i}\; \mu_{r'} \; \frac{1}{C^3_{r'}}    \left\{ \begin{array}{ccc} 1 & 2 & r' \\  3 & 4 & r \end{array} \right\}  \; \left\{ \begin{array}{ccc} 1 & 4 & v_3 \\ \tilde{v}_0 & \tilde{v}_2 & \tilde{r}' \end{array} \right\} \; \left\{ \begin{array}{ccc} 3 & 2 & \tilde{v}_1 \\  v_2 & v_0 & r' \end{array} \right\}    \\
     = \frac{1}{C_{r}}  \left\{ \begin{array}{ccc} 3 & 4 & \tilde{v}_0 \\  v_3 & v_1 & \tilde{r} \end{array} \right\} \; \left\{ \begin{array}{ccc} 1 & 2 & v_2 \\  \tilde{v}_1 & \tilde{v}_3 & r \end{array} \right\}          .   
\eea{pentaspecial2}

Let us now consider the crossing equation in \eqref{posind} and redefine the OPE functions as $K_{ijk} \rightarrow \frac{1}{C_i C_j C_k \sqrt{m_{i,ijk} m_{j,ijk} m_{k,ijk}}}\; K_{ijk}$\footnote{We'd like to remind the reader that when $O_i$ is a scalar, $C_i$ is equal to a constant $C$, and also that the labels corresponding to the scalars in $m_{i,ijk}$ are dropped.}. Using $n_r = \mu_r m_{r,r}^2$ (as can be easily shown using \eqref{cpwexp1}, \eqref{ortho}, and \eqref{bubble}) now gives us the crossing equation as,
\beq
\frac{1}{C_{r}} K_{12r}K_{34\tilde{r}}  = \sum_{\ell_{r'}} \int_{\frac{d}{2}}^{\frac{d}{2}+i\infty} \frac{d\Delta'}{2\pi i}\; \mu_{r'} \; \frac{1}{C_{r'}^{3}} K_{32r'}K_{14\tilde{r'}}  \; \left\{ \begin{array}{ccc} 1 & 2 & r' \\  3 & 4 & r \end{array} \right\} ,  \label{posindredef}
\eeq
which almost looks like the pentagon identity in \eqref{pentaspecial2}.

Comparing \eqref{pentaspecial2} and \eqref{posindredef} we see that setting $v_2 = \tilde{v}_0 = \tilde{v}_1 = v_3 = v$  let us set the redefined OPE coefficients equal to a one parameter ($v$) family of 6j symbols
\beq
K_{ijk} \overset{?}{=} \left\{ \begin{array}{ccc} i & j & \tilde{v} \\  \tilde{v} & v & k \end{array} \right\}   \quad,    \label{oneparam}
\eeq
such that they satisfy the position idependent crossing equation \eqref{posindredef}.

Let us now check if the OPE function in \eqref{oneparam} is consistent with the permutation symmetries of a bosonic correlator, see App. \ref{sec:tsconv}. For example, the OPE function $K_{ijk}$ should be symmetric under the swap (an acyclic permutation) $i\leftrightarrow j$. The 6j symbol on the r.h.s. of \eqref{oneparam} however does not in general have this property as it is clear from \eqref{tetra} (using the equality between the first and the fourth 6j symbol). To satisfy this amounts us to set $\tilde{v} =v = \frac{d}{2}$, and therefore we now have,
\beq
K_{ijk} = \left\{ \begin{array}{ccc} i & j & \frac{d}{2} \\  \frac{d}{2} & \frac{d}{2} & k \end{array} \right\}  \quad \quad  .    \label{6jsoln}
\eeq 
The tetrahedral symmetry equating the first and the third 6j symbol in \eqref{tetra} ensures that $K_{ijk}$ as defined in \eqref{6jsoln} is symmetric under a cyclic permutation to $K_{jki}$. All cyclic and acyclic permutations of $\{i,j,k\}$ can be generated from the two permutations just discussed and thus we see that the OPE function in \eqref{6jsoln} has all the desired permutation symmetries. 

The unitarity property in \eqref{unitarity1} is manifestly satisfied when $O_k$ is on the principal series as it is just the statement that the product of \eqref{6jsoln} and its complex conjugate is positive. Thus we have in \eqref{6jsoln} (with appropriate pre-factors as mentioned above) a solution to the bootstrap problem in sec. \ref{sec:reps}. %This is the main result of this paper, and it gives us an exact solution to the cosmological bootstrap as discussed in sec. \ref{sec:cosmoboot}. 

%Let us also take a moment to consider the unitarity of the solution in the context of a Lorentzian theory. In this case we'd want the solution $K_{ijk}$ to satisfy reflection positivity which can be stated in terms of the OPE coefficients as $\lambda_{ijO} = \lambda^{*}_{jiO}$ for the exchange of an operator $O$ with a real valued dimension. For oeprators with real dimensions, the 6j symbol in \eqref{6jdef} is an integral over a real valued function and is therefore real valued. The OPE coefficients are the residues of $K_{12r}$ (and similarly of $K_{34\tilde{r}}$) in $\Delta_r$. If the spectrum is real valued, then the residues are real and symmetric in $i\leftrightarrow j$ and reflection positivity is satisfied. However, we shall not be able to discuss this in explicit detail in the present work.

%%%%%%%%%%%%%%%%%%%%%%%%%%%%%%%%%%%%%%%%%%%%%%%%%%%%%%%%%%%%%%%%%%%%%%%%%%%%%%%%%%%%%%%%%%
\subsection{The solution in two dimensions}
\label{sec:2d}
%%%%%%%%%%%%%%%%%%%%%%%%%%%%%%%%%%%%%%%%%%%%%%%%%%%%%%%%%%%%%%%%%%%%%%%%%%%%%%%%%%%%%%%%%%

In this section we discuss the 6j symbol solution to the bootstrap problem a bit more explicitly. This is however complicated by the fact that the expression for a $SO(d+1,1)$ 6j symbol is known in closed form in only two and four dimensions \cite{Liu:2018jhs}\footnote{See also \cite{Isma, Derkachov:2017mip}.} and that these expressions are rather complicated and hence difficult to analyse rigorously. Therefore we shall restrict ourselves to a brief discussion of the two dimensional case. 

We have quoted the general expression for the 6j symbol with all spinning operators in two dimensions from \cite{Liu:2018jhs} in Appendix. \ref{sec:6j2dexp}. Let us now discuss how we can use this general expression and specialise to the 6j symbol solution discussed in the previous section. We can consider the following 6j symbol in particular\footnote{We are now using the quantum numbers of the operators to label their respective positions in the 6j symbols.},
\beq
\left\{ \begin{array}{ccc} \Delta_1 & \Delta_2 & 1 \\  1 & 1 & [\Delta, \ell] \end{array} \right\}      \quad   .   \label{6jsp}
\eeq

If we plug in the appropriate values of the labels from \eqref{6jsp} into the expression in \eqref{6j2d1}\footnote{See App. \ref{sec:6j2dexp} for the notation being used here.} we see that it blows up as the factors of $K_{h',\bar{h}'}^{h_i , h_j , \bar{h}_i , \bar{h}_j }$ in each of the terms in \eqref{6j2d1} have a factor of $\Gamma(1-2h')$ or $\Gamma(1-2\tilde{h'})$ wherein we  need to use $h' = \frac{d}{4} = \frac{1}{2}$. However, this singularity is only superficial and the problem can be resolved in the following manner. Let us  take $\Delta'\rightarrow \Delta'_{\epsilon}=\frac{d}{2}+\epsilon$ in place of $\frac{d}{2}$ in \eqref{6jsp}. This can be thought of as working in $d+2\epsilon$ dimensions and we eventually take the $\epsilon\rightarrow 0$ limit. Therefore we want to calculate\footnote{Since $\ell'=0$ and $\ell_i = 0$ we have simplified the notation from $K_{h',\bar{h}'}^{h_i , h_j , \bar{h}_i , \bar{h}_j } $ to $K_{h'}^{h_i , h_j }$ while using \eqref{6j2d1} in \eqref{6jsp}. Similarly, the notation for $\mathcal{B}_{[h,\bar{h}],[h',\bar{h}']}^{h_i , \bar{h}_i }$ has been simplified to $\mathcal{B}_{[h,\bar{h}],[h']}^{h_i}$.},
\beq
\left\{ \begin{array}{ccc} \Delta_1 & \Delta_2 & 1+\epsilon \\  1 & 1 & [\Delta, \ell] \end{array} \right\}  = K_{\frac{1-\epsilon}{2}}^{\frac{\Delta_1}{2} , \frac{1}{2}} \;\; \mathcal{B}_{[h,\bar{h}],[\frac{1+\epsilon}{2}]}^{\frac{\Delta_1}{2}, \frac{\Delta_2}{2}, \frac{1}{2}, \frac{1}{2}  }  +   K_{\frac{1+\epsilon}{2}}^{\frac{1}{2} , \frac{\Delta_2}{2}} \;\; \mathcal{B}_{[h,\bar{h}],[\frac{1-\epsilon}{2}]}^{\frac{\Delta_1}{2}, \frac{\Delta_2}{2}, \frac{1}{2}, \frac{1}{2} }      \quad .     \label{6jepsilon}
\eeq
Say we expand $K_{\frac{1\pm \epsilon}{2}}^{\frac{\Delta}{2} , \frac{1}{2}}$ in $\epsilon$ as follows\footnote{$K^{a,b}_{\frac{1\pm\epsilon}{2}}$ is symmetric under $a \leftrightarrow b$.},
\beq
K_{\frac{1\pm \epsilon}{2}}^{\frac{\Delta}{2} , \frac{1}{2}} = \frac{f^{(0)}_{\Delta,\pm}}{\epsilon} + f^{(1)}_{\Delta,\pm} +\mathcal{O}\left(\epsilon\right)      ,          \label{Kexp}
\eeq
and $\mathcal{B}_{[h,\bar{h}],[\frac{1\pm\epsilon}{2}]}^{\frac{\Delta_1}{2}, \frac{\Delta_2}{2}, \frac{1}{2}, \frac{1}{2} }$ as,
\beq
\mathcal{B}_{[h,\bar{h}],[\frac{1\pm\epsilon}{2}]}^{\frac{\Delta_1}{2}, \frac{\Delta_2}{2}, \frac{1}{2}, \frac{1}{2} } =  \mathcal{B}_{[h,\bar{h}],[\frac{1}{2}]}^{\frac{\Delta_1}{2}, \frac{\Delta_2}{2}, \frac{1}{2}, \frac{1}{2} }    +  \epsilon \left(\mathcal{B}_{\epsilon,\pm}\right)_{[h,\bar{h}],[\frac{1}{2}]}^{\frac{\Delta_1}{2}, \frac{\Delta_2}{2}, \frac{1}{2}, \frac{1}{2} }     +   \mathcal{O}\left(\epsilon^2\right)      \quad . \label{Bexp}
\eeq
$\mathcal{B}_{[h,\bar{h}],[\frac{1\pm\epsilon}{2}]}^{\frac{\Delta_1}{2}, \frac{\Delta_2}{2}, \frac{1}{2}, \frac{1}{2} }$ is regular as $\epsilon \rightarrow 0$, and $\left(\mathcal{B}_{\epsilon,\pm}\right)_{[h,\bar{h}],[\frac{1}{2}]}^{\frac{\Delta_1}{2}, \frac{\Delta_2}{2}, \frac{1}{2}, \frac{1}{2} }$ is its derivative w.r.t. $\epsilon$ evaluated at $\epsilon = 0$. Thus we obtain,
\beq
\left\{ \begin{array}{ccc} \Delta_1 & \Delta_2 & 1 \\  1 & 1 & [\Delta,\ell] \end{array} \right\} = \mathcal{B}_{[h,\bar{h}],[\frac{1}{2}]}^{\frac{\Delta_1}{2}, \frac{\Delta_2}{2}, \frac{1}{2}, \frac{1}{2} } \left(f^{(1)}_{\Delta_2 ,+} + f^{(1)}_{\Delta_1 ,-}\right)  
			+  f^{(0)}_{\Delta_1 ,-} \left(\mathcal{B}_{\epsilon,+}\right)_{[h,\bar{h}],[\frac{1}{2}]}^{\frac{\Delta_1}{2}, \frac{\Delta_2}{2}, \frac{1}{2}, \frac{1}{2} }  
		 			+  f^{(0)}_{\Delta_2 ,+} \left(\mathcal{B}_{\epsilon,-}\right)_{[h,\bar{h}],[\frac{1}{2}]}^{\frac{\Delta_1}{2}, \frac{\Delta_2}{2}, \frac{1}{2}, \frac{1}{2} }  . \label{6j2dsimp}
\eeq
The important simplification here is that the term proportional to $\epsilon^{-1}$ vanishes as $f^{(0)}_{\Delta,+} $ is equal to $- f^{(0)}_{\Delta,-}$ and is independent of $\Delta$.

We can evaluate \eqref{6j2dsimp} using the results quoted in Appendix  \ref{sec:6j2dexp}. The resulting expression is extremely long and complicated, so we do not present it here. As a consistency check we can verify numerically that the expression satisfies the following (residual) tetrahedral symmetries\footnote{Note that the general expressions for the 6j symbol in Appendix. \ref{sec:6j2dexp} are not manifestly tetrahedrally symmetric although the presence of the symmetries can be verified numerically.} which also serves as a consistency check that the 6j symbol solution for the three-point function coefficient has all the requisite permutation symmetries,
\beq
\left\{ \begin{array}{ccc} \Delta_1 & \Delta_2 & 1 \\  1 & 1 & [\Delta,0] \end{array} \right\} = \left\{ \begin{array}{ccc} \Delta_2 & [\Delta,0] & 1 \\  1 & 1 & \Delta_1 \end{array} \right\} = \left\{ \begin{array}{ccc} \Delta_2 & \Delta_1 & 1 \\  1 & 1 & [\Delta,0] \end{array} \right\}     \quad .     \label{tetrahedralresi}
\eeq

A second consistency check for our results is the following condition of shadow symmetry followed by the OPE function $\rho(\Delta_R, \ell_R) = \mu(\Delta_R , \ell_R ) K_{12R}K_{34\tilde{R}}$\footnote{The factors of $C_i$ and $m_{i,ijk}$ involved in the rescaling of the three point function coefficients in sec. \ref{sec:crossoln} are shadow symmetric, hence they cancel out across the two sides of \eqref{symmetrysha}.},
\beq
\rho(\Delta_R,\ell_R) S(O_3 O_4 [\tilde{O}_R]) = \rho(\tilde{\Delta}_R,\ell_R) S(O_1 O_2 [\tilde{O}_R])     \quad  .       \label{symmetrysha}
\eeq
$S(O_i O_j [\tilde{O}])^{a}_{b}$ are matrix elements of a map to a different basis of tensor structures that is caused by a shadow transform\footnote{When the tensor structure is unique, as is the case for us, we can drop the indices from the shadow coefficients.}. We have defined these in App. \ref{sec:cpwcb}, and they have been calculated explicitly in certain cases in \cite{Karateev:2018oml}. 

As we show in App. \ref{sec:cpwcb}, the symmetry in \eqref{symmetrysha} is necessary to obtain a block expansion from the CPW expansion with the OPE function $\rho(\Delta_R,\ell_R)$ but this is always satisfied as \eqref{symmetrysha} is a consequence of the Euclidean Inversion Formula (EIF) \cite{Caron_Huot_2017, Kravchuk:2018htv}. We have used the expression in \eqref{6j2dsimp} to obtain $\rho(\Delta_R,\ell_R)$ and the expressions for the shadow coefficients from \cite{Karateev:2018oml} to perform numerical checks that our solution satisfies the shadow symmetry in \eqref{symmetrysha}. This serves as a verification that the solution is indeed consistent with the EIF applied on a conformally covariant correlation function.

The contour integral in the CPW expansion picks out the poles in $\rho(\Delta_R,\ell_R) S(O_3 O_4 [\tilde{O}_R])$\footnote{The integration contour is typically chosen/deformed to pick up the series of poles extending to the right and avoid the series extending to the left.} to give us the conformal block expansion. There are a few different series of poles in this function that can be identified from the expression. The series that extend to the right are as follows:
\bea
\Delta_R & =  2 - \ell_R - \Delta_1 + 2n   \quad ,\\
\Delta_R & =  \ell_R + \Delta_1 + 2n   \quad ,\\
\Delta_R & =  2 - \ell_R - \Delta_3 + 2n   \quad ,\\
\Delta_R & =  \ell_R + \Delta_3 + 2n   \quad ,\\
\Delta_R & =  \ell_R + 1 + n      \quad,      \\
\Delta_R & =  \ell_R + 2 + 2n   \quad \text{(double)} \quad ,  \\    
\Delta_R & =  \Delta_1 +\Delta_2 + \ell_R + 2n   \quad , \\
\Delta_R & =  \Delta_3 +\Delta_4 + \ell_R + 2n   \quad ,
\eea{poles2d}  
with $n$ being an integer $\geq 0$. All poles are simple poles unless explicitly stated otherwise. Consider the first four series of poles listed above in \eqref{poles2d}. These are not compatible with the tetrahedral symmetry that we expect in our 6j symbols. We can indeed verfiy numerically, that although there are gamma functions in \eqref{6j2dsimp} that give rise to these poles, they are actually cancelled out across the different terms in the expression. 

The poles at $\Delta_R =  \ell_R + 1 + k $ are simple poles in the shadow coefficient $S(O_3 O_4 [\tilde{O}_R])$. The pole at $k=0$ is cancelled by a corresponding zero in the factor $\mu(\Delta_R , \ell_R )$, so we are left with poles for $k\geq 1$. In the generic case when there are no overlapping poles from $K_{12R}K_{34\tilde{R}}$, as shown in \cite{Simmons_Duffin_2018}, these poles from $S(O_3 O_4 [\tilde{O}_R])$ cancel with a set of poles at $\Delta_R = \ell_R+d-1-k$ for $k=1,\cdots,J$ in the conformal blocks themselves. In this case however, for odd values of $k$, we have an overlap with the next series of (double) poles at $\Delta_R =  \ell_R + 2 + 2n $ for $n \geq 0$. The contribution from the poles in the blocks can still be cancelled but we are left with part of the residues from the triple poles at $\Delta_R =  \ell_R + 2 + 2n$. We don't have a clear interpretation for these poles yet and we leave it to future work.  

The next two sets of poles are interesting as they are reminiscent of exchanged double trace operators in tree level Witten diagrams in AdS. These are generically simple poles but for pairwise equal external operators, we get double poles at these double trace operators. It has been shown in \cite{Sleight:2020obc,Sleight:2021plv} that exchange diagrams in dS can be obtained as linear combinations of exchange Witten diagrams in Euclidean AdS. Note that \eqref{posindredef} is an equality between the CPW expansions in the s-channel and the t-channel. We can swap $3\leftrightarrow 4$ in this equation to obtain the corresponding equality between the s-channel and the u-channel expansions and this will verify that the solution is of the form in \eqref{6jsoln} in all channels. Thus we see that the solution we have obtained features double-trace poles of unbounded spin in all channels. This pole structure suggests a sum over exchange Witten diagrams in all three channels. However, the 6j symbol solution which is analytic in $\ell_R$, can only capture the entirety of OPE data for $\ell_R$ greater than the spin of the specific operator being exchanged. The double (and single) trace data in the channel of the exchange is non-analytic as the spin of the double trace operators is bounded above by the spin of the exchanged operator. Unfortunately, we cannot provide a more complete and satisfactory analysis/interpretation of the pole structure just yet and we would like to address these questions in future work. 

%We can calculate the residue of the OPE function $C(\Delta_R, \ell_R)$, see \eqref{blockint}, at the double trace poles, and extract the OPE coefficients from the pure power terms (ignoring the $\log u$ terms in the block expansion that would correspond to the anomalous dimension). We have done this numerically on some examples and observed that the OPE coefficients are not of a definite sign. Therefore the solution does not correspond to that of a unitary CFT. 

We have outlined how to obtain the explicit expression for the 6j symbol solution to the crossing equation in two dimensions and touched upon its analytic structure. This is as far as we'll go in discussing concrete examples for now. The expression for the 6j symbol is also known in four dimensions and a preliminary analysis should be, in principle, possible in this case as well. However the solution is even more complicated and tedious in this case and we shall not delve into this matter in this paper.

%%%%%%%%%%%%%%%%%%%%%%%%%%%%%%%%%%%%%%%%%%%%%%%%%%%%%%%%%%%%%%%%%%%%%%%%%%%%%%%%%%%%%%%%%%
\section{Conclusions}
\label{sec:conc}
%%%%%%%%%%%%%%%%%%%%%%%%%%%%%%%%%%%%%%%%%%%%%%%%%%%%%%%%%%%%%%%%%%%%%%%%%%%%%%%%%%%%%%%%%%

The QFT living on the late-time boundary of dS space is covariant under the dS isometry group $SO(d+1,1)$ and the boundary correlators evaluated in a Bunch-Davies vacuum form a Euclidean CFT. However, this is not a unitary theory in the Lorentzian sense and the spectrum generically consists of operators with dimensions taking complex values. It is interesting to calculate the late time values of correlation functions in dS to study  the evolution of the universe at cosmological scales, and these can be approximated by the conformal correlation functions on the late time boundary. 

Recently, there has been an interest in importing ideas from the non-perturbative conformal bootstrap programme \cite{Poland:2022qrs,Hartman:2022zik} to the study of cosmological correlation functions \cite{Baumann:2022jpr}, particularly in the context of the CFT on the late-time boundary  \cite{Hogervorst:2021uvp}. The main tools here are the crossing equations between the conformal partial wave expansions in different channels and the positivity of the spectral function for unitary irreps. In this paper, our main goal has been to construct an exact solution to this crossing equation \eqref{stcrossing} for cosmological bootstrap. 

We started with an overview of conformal representation theory leading to the expansion of a scalar four point function in terms of CPWs on the principal series. The (almost) completeness of this basis allows us to equate the analogous expansions in different channels thus giving us the crossing equation we want to solve. The position dependence of this crossing equation is superficial at best as we can use the orthogonality of the CPWs to rewrite it simply in terms of the three-point function coefficients with a kernel made of a 6j symbol \eqref{posind}.

A position independent crossing equation for a generic group was suggested in \cite{Gadde:2017sjg} and a class of solutions using 6j symbols is presented for the special case of SU(2). The crucial ingredient here is a pentagon identity with the 6j symbols of the group, also known as the Biedenharn-Elliot identity. In this paper, we have presented a detailed derivation for the analogous pentagon identity for 6j symbols of the Euclidean conformal group, making use of some modern tools that have resulted from the recent resurgence of interest in the harmonic analysis on the conformal group \cite{Karateev:2018oml,Kravchuk:2016qvl}. This pentagon identity can be morphed into the position independent crossing equation mentioned above when the parameters are chosen appropriately. This gives an exact solution to the crossing equation where the three-point function coefficients are given in terms of certain 6j symbols \eqref{6jsoln}. This 6j symbol solution to the cosmological bootstrap equations is the main result of this paper. 

These three-point function coefficients form the spectral function in the CPW expansion of the correlators, and carry the dynamical information of the boundary theory. Therefore an immediate follow-up question after the general derivation is to analyse the analytical structure of the spectral function in specific examples. We discuss this briefly in the case of two spacetime dimensions using the expressions for the 6j symbols derived in \cite{Liu_2019}. The highlight here is the presence of simple poles corresponding to double-trace like operators which coalesce to double poles for pairwise equal external operators. This is reminiscent of tree level Witten diagrams, which is particularly interesting given the connection between dS scattering and Witten diagrams in EAdS \cite{Sleight:2021plv,Sleight:2020obc,DiPietro:2021sjt}. The difference from a specific Witten diagram lies in that the spectral function is analytic in spin in all channels. Additionally there is a class of poles independent of the dimensions of external operators that we cannot explain at the moment. 

In this paper, we have only attempted to study the pole structure in the case of two boundary dimensions for simplicity, and it would be good to look at the case of four dimensions too to see whether it exhibits similar features and difficulties. The 6j symbols in four dimensions are known in closed form \cite{Liu_2019}. 

We would like to investigate in the future the possibility of connecting our solution to exchange diagrams in EAdS, and provide a more complete analysis of the pole structure of the spectral function. In general, it would be interesting to find an interpretation of this solution in terms of a known CFT, holographic, or otherwise. To that end, it might be useful to extend our results to incorporate spinning external operators which may reveal a larger class of 6j symbol solutions. In particular, it's natural to wonder if a pentagon identity and consequently, a solution to the crossing equation can be obtained using the 6j symbols for fermions \cite{Albayrak:2020rxh}. 

A satisfactory understanding of the CFT on the late-time boundary would let us ask what it implies for physics in the bulk dS spacetime. We also hope that this work provides additional evidence that 6j symbols are important objects in the study of CFTs, and it would be nice to have closed form expressions or a better handle on the analytic structure of 6j symbols in general dimensions and for quantum numbers with non-zero spin.  

It would also be interesting to explore the case of broken conformal symmetry, especially in the presence of defects and boundaries. There has been significant interest in incorporating these non-localities in conformal bootstrap using both analytical and numerical methods \cite{Liendo:2012hy,Gliozzi:2015qsa,Billo:2016cpy,Lemos:2017vnx,Lauria:2018klo}, and also in the context of the AdS/CFT correspondence \cite{Maldacena:1997re,Witten:1998qj,deLeeuw:2017dkd}\footnote{We do not provide an exhaustive list of references on defect CFT here as this is beyond the scope of this paper, one can look at \cite{Andrei:2018die} (for example) and the references therein for a better survey of the relevant literature.}. This would also involve developing the relevant technology when the symmetry group in question is generically the conformal group times a compact group.

%!TEX root = ../Central_compile.tex
%%%%%%%%%%%%%%%%%%%%%%%%%%%%%%%%%%%%%%%%%%%%%

\section{Acknowledgements}
\label{sec:ack}

We would like to thank Ant\'{o}nio Antunes, Luca Cassia, Lorenzo Di Pietro, Simon Ekhammar, Giulia Fardelli, Henrik Johansson, Aaditya Salgarkar, Oliver Schlotterer, Jacopo Sisti, and Charles Thull, for relevant discussions along the course of the project, and Jacopo Sisti and Lorenzo Di Pietro for their comments on the draft. SS is especially thankful to Andrea Manenti for many fruitful discussions directly relevant to this work and for his comments on the draft, to Kamran Salehi Vaziri for a detailed discussion on the work, and to Martijn Hidding for his help with speeding up numerical evaluations.  This work is partially funded by the Knut and Alice Wallenberg Foundation grant KAW 2021.0170, VR grant 2018-04438 and Olle Engkvists Stiftelse grant 2180108.

\appendix

%%%%%%%%%%%%%%%%%%%%%%%%%%%%%%%%%%%%%%%%%%%%%%%%%%%%%%%%%%%%%%%%%%%%%%%%%%%%%%%%%
\section{Tensor structure conventions}
\label{sec:tsconv}
%%%%%%%%%%%%%%%%%%%%%%%%%%%%%%%%%%%%%%%%%%%%%%%%%%%%%%%%%%%%%%%%%%%%%%%%%%%%%%%%%

In this section, we want to outline the idea behind the conventions we follow for the tensor structures and the OPE coefficients. The exact choice of conventions for specific correlators is both tedious to state and irrelevant since only the relative signs matter. So it should suffice as long as we follow the principles stated below in a consistent manner.

Let us first consider the simplest example of a scalar-scalar-vector correlator. The only tensor structure in this case is as follows,
\beq
\braket{O_1 (x) O_2 (y) O^{\mu} (z)} = V_{12,O}^{\mu} =\frac{(x-z)^{\mu}}{|x-z|} - \frac{(y-z)^{\mu}}{|x-z|}      \quad .   \label{tens1}
\eeq
The tensor structure $V_{ij,O}^{\mu}$ is defined such that the sign of the position at which $O_i$ is inserted has positive sign. In this case we have chosen $V_{12,O}^{\mu}$ to be the tensor structure. Note that we could have equally well chosen to work with $V_{21,O}^{\mu}$ in which case the position of $O_2$ would have positive sign instead as $V_{12,O}^{\mu} = -V_{21,O}^{\mu}$. The choice of convention for the tensor structures thus involves fixing the sign of the position at which specific operators are inserted. For example, if we insert the operators above at different positions, say $x_{3}$, $x_4$, and $x$ respectively, the tensor structure consistent with \eqref{tens1} is as follows,
\beq
\braket{O_1 (x_3) O_2 (x_4) O^{\mu} (x)} = V_{12,O}^{\mu} = \frac{(x_3 -x)^{\mu}}{|x_3 -x|} - \frac{(x_4 -x)^{\mu}}{|x_4 -x|}      \quad .   \label{tens2}
\eeq 

If we consider a correlator of say two vectors and a scalar, there's one more tensor structure in addition to the one already mentioned,
\beq
\braket{O_1^{\mu_1} (x) O_2^{\mu_2} (y) O (z)} = H_{12}^{\mu_1 \mu_2} = \eta^{\mu_1 \mu_2} - 2\frac{(x-y)^{\mu_1}(x-y)^{\mu_2}}{(x-y)^{2}}     \quad.   \label{tens3}
\eeq
In this case, there's no sign ambiguity involved as $H_{12}^{\mu_1 \mu_2} = H_{21}^{\mu_1 \mu_2}$. All tensor structures in any correlator of bosonic operators can be built out of these basic building blocks in \eqref{tens1} and \eqref{tens3} \cite{Costa:2011mg}.

For the 3-pt function coefficients, the ordering of the labels are specified using the positions at which the operators are inserted. For example, the coefficients for the three point function $\braket{\braket{O_i (x_1) O_j (x_2) O_k (x_3)}}$\footnote{We have used $\braket{\braket{\cdots}}$ for a correlator to distinguish it from a tensor structure that we denote as $\braket{\cdots}$.} can be denoted as $K_{ijk}$. Here we have chosen the ordering for the labels of the 3-pt function coefficient to be that of the operators at $x_1$, $x_2$, $x_3$ respectively. So the 3-pt fn coefficient for $\braket{\braket{J^{\mu} (x_2) O_1 (x_1) O_2 (x_3)}}$, $J^{\mu}$ being a vector current, is $K_{O_1 J O_2}$. Note that if there are more than one tensor structure in the correlator, then we have a 3-pt function coefficient associated with each of them that we can choose to differentiate notationally with a superscript.

For bosonic correlators in Euclidean spacetime, we can freely swap the order of the operators in the correlator, that is $\braket{\braket{J^{\mu} (x_2) O_1 (x_1) O_2 (x_3)}} = \braket{\braket{J^{\mu} (x_2)  O_2 (x_3) O_1 (x_1)}}$ (for example). However exchanging the positions at which the operators are inserted can change things. Consider the correlator $\braket{\braket{J^{\mu} (x_2) O_1 (x_3) O_2 (x_1)}}$. The associated tensor structure, defined in \eqref{tens1}, picks up a sign. The correlator itself picks up this overall sign while the 3-pt fn coefficient is symmetric under the swap $K_{O_1 J O_2} = K_{O_2 J O_1}$. Just to reiterate a point that we have already made: we could have chosen to denote the coefficient of $\braket{\braket{O_i (x_1) O_j (x_2) O_k (x_3)}}$ as $K_{ikj}$ just as well, in which case the coefficient in $\braket{\braket{J^{\mu} (x_2) O_1 (x_1) O_2 (x_3)}}$ would be denoted as $K_{O_1 O_2 J}$. We only need to be consistent in the choices we make.      

%%%%%%%%%%%%%%%%%%%%%%%%%%%%%%%%%%%%%%%%%%%%%%%%%%%%%%%%%%%%%%%%%%%%%%%%%%%%%%%%%
\section{6j symbols in two dimensions}
\label{sec:6j2dexp}
%%%%%%%%%%%%%%%%%%%%%%%%%%%%%%%%%%%%%%%%%%%%%%%%%%%%%%%%%%%%%%%%%%%%%%%%%%%%%%%%%

In this section we state the general expression for the 6j symbol with all spinning operators in two dimensions that we use in sec. \ref{sec:2d}. We have taken this result directly from \cite{Liu:2018jhs} where it has been calculated using the Lorentzian inversion formula \cite{Caron_Huot_2017, Simmons_Duffin_2018, Kravchuk:2018htv} exploiting the fact that the 6j symbol is the ``OPE function" in the expansion of a CPW in one channel in terms of those in a different channel as shown in \eqref{cpwexp} and \eqref{cpwpair}. We shall borrow the notation from the source article: $h_i = \frac{\Delta_i + \ell_i}{2}$, $\bar{h}_i = \frac{\Delta_i - \ell_i}{2}$, $\tilde{h}_i = 1-h_i$, and $\tilde{\bar{h}}_i = 1- \bar{h}$. 
\beq
\left\{ \begin{array}{ccc} O_1 & O_2 & O' \\  O_3 & O_4 & O \end{array} \right\} = K_{\tilde{h}',\tilde{\bar{h}}'}^{h_1 , h_4 , \bar{h}_1 , \bar{h}_4 } \mathcal{B}_{[h,\bar{h}],[h',\bar{h}']}^{h_i , \bar{h}_i }  +   K_{h',\bar{h}'}^{h_3 , h_2 , \bar{h}_3 , \bar{h}_2 } \mathcal{B}_{[h,\bar{h}],[\tilde{h}',\tilde{\bar{h}'}]}^{h_i , \bar{h}_i }      \quad  ,  \label{6j2d1}
\eeq
where,
\beq
 K_{h',\bar{h}'}^{h_i , h_j , \bar{h}_i , \bar{h}_j } = \frac{\pi \Gamma\left(1-2\tilde{h}\right)\Gamma\left(\tilde{h}-h_{ij}\right)\Gamma\left(\tilde{\bar{h}}+\bar{h}_{ij}\right)}{\Gamma\left(2\tilde{\bar{h}}\right)\Gamma\left(h-h_{ij}\right)\Gamma\left(\bar{h}+\bar{h}_{ij}\right)}  \quad   , \quad h_{ij} = h_i - h_j  \quad,     \label{6j2d2}
\eeq
and 
\bea
\mathcal{B}_{[h,\bar{h}],[h',\bar{h}']}^{h_i , \bar{h}_i } = & \quad \frac{(-1)^{j_H}}{4}\; \frac{\Gamma\left(h+h_{12}\right) \Gamma\left(h+h_{21}\right) \Gamma\left(\tilde{\bar{h}}+\bar{h}_{43}\right)\Gamma\left(\tilde{\bar{h}}+\bar{h}_{34}\right)}{\Gamma(2h)\Gamma(2-2\bar{h})}     \\
			& \sin\left(\pi\left(\bar{h}'-\bar{h}_1 -\bar{h}_4 \right)\right) \; \sin\left(\pi\left(\bar{h}'-\bar{h}_2 - \bar{h}_3 \right)\right)\; \Omega_{h,h',h_2 +h_3}^{h_i } \Omega_{\tilde{\bar{h}},\bar{h}', \bar{h}_2 +\bar{h}_3}^{\bar{h}_i }  \quad .
\eea{6j2d3}
$j_H$ is short for $h- \bar{h} -\tilde{h}_3 + \bar{\tilde{h}}_3 -\tilde{h}_4 + \bar{\tilde{h}}_4$, and $\Omega_{h,h',p}^{h_i }$ is as follows,
\bea
\Omega_{h,h',p}^{h_i } \; = & \quad \frac{\Gamma(2h)\Gamma(h' -p+1)\Gamma(h'-h_{12}+h_{34}-p+1)\Gamma(-h'+h_{12}+h+p-1)}{\Gamma(h_{12}+h)\Gamma(h_{34}+h)\Gamma(h'-h_{12}+h-p+1)}     \\
				& \quad \pFq{4}{3}{h'+h_{23},h'-h_{14},h'-h_{12}+h_{34}-p+1,h'-p+1}{2h',h'-h_{12}+h-p+1,h'-h_{12}-h-p+2}{1}  \\
				& \;\; + \frac{\Gamma(2h')\Gamma(h'-h_{12}-h-p+1)\Gamma(h_{13}+h+p-1)\Gamma(h_{42}+h+p-1)}{\Gamma(h'+h_{23})\Gamma(h'-h_{14})\Gamma(h'+h_{12}+h+p-1)}  \\
				& \quad \quad \pFq{4}{3}{h_{13}+h+p-1, h_{42}+h+p-1, h_{34}+h, h_{12}+h}{h' +h_{12}+h+p-1, 2h, -h' +h_{12}+h+p}{1}    \quad .
\eea{6j2d4}

%\input{sections/Appendix_6j.tex}
%%%%%%%%%%%%%%%%%%%%%%%%%%%%%%%%%%%%%%%%%%%%%%%%%%%%%%%%%%%%%%%%%%%%%%%%%%%%%%%%%
\section{Conformal partial waves to conformal blocks}
\label{sec:cpwcb}
%%%%%%%%%%%%%%%%%%%%%%%%%%%%%%%%%%%%%%%%%%%%%%%%%%%%%%%%%%%%%%%%%%%%%%%%%%%%%%%%%

We shall now quickly recap how we can obtain a conformal block expansion from a CPW expansion. We start from the CPW expansion in \eqref{cpwexp2},
\beq
\braket{O_1 (x_1) O_2 (x_2) O_3 (x_3) O_4 (x_4)} =  \sum_{\ell_R } \int_{\frac{d}{2}}^{\frac{d}{2}+i\infty} \frac{d\Delta_R }{2\pi i} \; \rho(\Delta_R , \ell_R ) \; \Psi^{R}_{1234}(x_i )  \quad, \label{cpwapp}
\eeq
where we have used $\rho(\Delta_R , \ell_R ) = \mu(\Delta_R,\ell_R) \; K_{12R}K_{34\tilde{R}}$. The Plancherel measure $\mu(\Delta_R,\ell_R)$ is a measure defined on the space of unitary irreps of a group $G$. For the Euclidean conformal group (which is non-compact), the Plancherel measure is given by \cite{Karateev:2018oml},
\beq
\frac{\mu(\Delta,\sigma)}{\text{Vol } SO(1,1)} = \frac{\text{Tr}_{\Delta,\sigma} (1)}{\text{Vol }SO(d+1,1)}   \quad.           \label{planch}
\eeq
The volume of the group is calculated w.r.t. the Haar measure. Both sides of \eqref{planch} are defined formally so as to define a finite Plancherel measure. The Plancherel measure is known in several cases and we refer the interested reader to \cite{Karateev:2018oml, Dobrev:1977qv} for further details inc. explicit expressions. 

In order to obtain the conformal block expansion from \eqref{cpwapp}, let us first express the CPWs in terms of blocks $G^{R}_{1234}$ \cite{Karateev:2018oml}:
\beq
\Psi^{R}_{1234}  = S(O_3 O_4 [\tilde{O}_R]) G^{R}_{1234} + S(O_1 O_2 [O_R ]) G^{\tilde{R}}_{1234}     \quad.   \label{cpwtocb}
\eeq
Here we have introduced the shadow coefficients $S(\cdots)$, which in general are defined as follows \cite{Karateev:2018oml},
\beq
\braket{\mathbb{S}[O_1] O_2 O_3}^{(a)} = S\left([O_1 ] O_2 O_3 \right)^a_{b} \braket{\tilde{O}_1 O_2 O_3}^{(b)}     \quad, \label{shadowco}
\eeq
The labels in subscript and superscript range over different basis of tensor structures, and there is an implied summation over $b$. The labels are redundant when there is a unique tensor structure as is the case if two of the operators are scalars. $\mathbb{S}[O]$ is the shadow transform of $O$ defined as follows,
\beq
\mathbb{S}[O] (x) = \int_{\mathbb{R}^d} dy \braket{\tilde{O}(x)\tilde{O}(y)} \; O(y)      \quad.   \label{shadowtrans}
\eeq
Thus the $S(\cdots)^a_b$ coefficients form a matrix transforming to a different basis of tensor structures under a shadow transform. These coefficients (or matrix elements) have been calculated for certain cases in \cite{Karateev:2018oml}.

Assuming \eqref{symmetrysha} holds true, we can rewrite \eqref{cpwapp} as an integral over an extended contour as follows,
\beq
\braket{O_1 (x_1) O_2 (x_2) O_3 (x_3) O_4 (x_4)} =  \sum_{\ell_R } \int_{\frac{d}{2}-\infty}^{\frac{d}{2}+i\infty} \frac{d\Delta_R }{2\pi i} \; C(\Delta_R , \ell_R ) \; G^{R}_{1234}(x_i )  \quad,    \label{blockint}
\eeq
where $C(\Delta_R , \ell_R ) = \rho(\Delta_R , \ell_R )\; S(O_3 O_4 [\tilde{O}_R])$. The conformal block $G^{R}_{1234}(x_i )$ decays exponentially on the right half $\Delta_R$ plane, therefore we can close the contour in \eqref{blockint} at positive real infinity. This picks up the poles in $C(\Delta_R , \ell_R )$ and we obtain the conformal block expansion. 

We shall now present a short proof of \eqref{symmetrysha} for the case when all external operators are scalars. To that end let's first quote the EIF in the language of shadow coefficients and three-point pairings \cite{Karateev:2018oml},
\beq
\frac{\rho(\Delta_R, \ell_R)}{\mu(\Delta_R,\ell_R)}\left(\braket{O_1 O_2 O}, \braket{\tilde{O}_1 \tilde{O}_2 \tilde{O}}\right) \left(\braket{O_3 O_4 \tilde{O}}, \braket{\tilde{O}_3 \tilde{O}_4 O}\right)  = \left(\braket{O_1 \cdots O_4}, \Psi^{\tilde{O}}_{\tilde{O}_{i}}\right)    \quad.      \label{EIFgen}
\eeq
The three-point function pairings in \eqref{EIFgen} have been calculated explicitly in \cite{Karateev:2018oml}. For our purposes it suffices to note that $\left(\braket{O_1 O_2 O}, \braket{\tilde{O}_1 \tilde{O}_2 \tilde{O}}\right)$ is a function of the spin of $O$ and has no dependence on operator dimensions. Let us denote this value by $f_{\ell_R}$. Therefore, we obtain from \eqref{cpwtocb} and \eqref{EIFgen},
\beq
\frac{\rho(\Delta_R, \ell_R)}{\mu(\Delta_R,\ell_R)}  = \frac{S(\tilde{O}_3 \tilde{O}_4 [O_R])\left(\braket{O_1 \cdots O_4}, G^{\tilde{O}}_{\tilde{O}_{i}}\right) + S(\tilde{O}_1 \tilde{O}_2 [\tilde{O}_R])\left(\braket{O_1 \cdots O_4}, G^{O}_{\tilde{O}_{i}}\right)}{f_{\ell_R}^2}    \quad.     \label{EIFgen2}
\eeq

The l.h.s. of \eqref{symmetrysha} (with an extra factor of the Plancherel measure in the denominator) $\frac{\rho(\Delta_R, \ell_R)}{\mu(\Delta_R,\ell_R)}S(O_3 O_4 [\tilde{O}_R])$ is thus given by,
\beq
\frac{S(\tilde{O}_3 \tilde{O}_4 [O_R])S(O_3 O_4 [\tilde{O}_R])\left(\braket{O_1 \cdots O_4}, G^{\tilde{O}}_{\tilde{O}_{i}}\right) + S(\tilde{O}_1 \tilde{O}_2 [\tilde{O}_R])S(O_3 O_4 [\tilde{O}_R])\left(\braket{O_1 \cdots O_4}, G^{O}_{\tilde{O}_{i}}\right)}{f_{\ell_R}^2}     \quad.   \label{EIFgen3}
\eeq
To proceed further, we use the fact that the shadow coefficient $S(O_3 O_4 [O_R])$ is symmetric under $\Delta_{3,4} \leftrightarrow \tilde{\Delta}_{3,4}$\footnote{See \cite{Karateev:2018oml} for the expression for $S(O_3 O_4 [O_R])$.}, and that $S(O_3 O_4 [O_R])S(O_3 O_4 [\tilde{O}_R])$ is the coefficient of a double shadow transform on $O_R$ in $\braket{O_3 O_4 O_R}$. Since the shadow of a shadow is proportional to the identity, this coefficient should just be a function of $\Delta_R$ and $\ell_R$. We won't be needing the explicit expression for the same and we shall just denote it as $N_R$. This gives us,
\beq
\frac{N_R\left(\braket{O_1 \cdots O_4}, G^{\tilde{O}}_{\tilde{O}_{i}}\right) + S(O_1 O_2 [\tilde{O}_R])S(O_3 O_4 [\tilde{O}_R])\left(\braket{O_1 \cdots O_4}, G^{O}_{\tilde{O}_{i}}\right)}{f_{\ell_R}^2}     \quad.     \label{EIFgen4}
\eeq

We can now consider the r.h.s. of \eqref{symmetrysha} and similar arguments can be used to show that $\frac{\rho(\tilde{\Delta}_R, \ell_R)}{\mu(\tilde{\Delta}_R,\ell_R)}S(O_1 O_2 [\tilde{O}_R])$ is also equal to \eqref{EIFgen4}. Since the Plancherel measure $\mu(\Delta_R,\ell_R)$ is shadow symmetric, this proves the shadow symmetry relation \eqref{symmetrysha} as a consequence of the EIF.

\bibliographystyle{JHEP}
\bibliography{draft6j}

\end{document}